\documentclass[dvipsnames]{bmcart}

\usepackage[utf8]{inputenc} 

\def\includegraphics{}

\usepackage[numbers, sort&compress]{natbib}
\newcommand{\ignore}[1]{}
\usepackage{fancyhdr}
\usepackage[normalem]{ulem}
\usepackage[hyphens]{url}
\usepackage{hyperref}
\usepackage{graphicx}
\usepackage{multirow}
\usepackage{color}
\usepackage{float}
\usepackage{enumitem}
\usepackage{amsmath}
\usepackage{arydshln}
\usepackage{booktabs}
\usepackage{xstring}
\usepackage[utf8]{inputenc}
\usepackage{gensymb}
\usepackage{pifont}
\usepackage{cleveref}
\usepackage{color,soul}
\usepackage{xstring}

\usepackage[most]{tcolorbox}
\tcbuselibrary{skins}
\usepackage{lipsum}

\usepackage{caption}
\usepackage{subcaption}
\usepackage{setspace}

\startlocaldefs
\endlocaldefs

\newif\ifcameraready
\camerareadytrue

\newcommand{\chI}[0]{}
\newcommand{\chII}[0]{}
\newcommand{\chIII}[0]{}
\newcommand{\chIV}[0]{}
\newcommand{\chV}[0]{}
\newcommand{\chVI}[0]{}
\ifcameraready
  \newcommand{\chVII}[0]{}
\else
  \newcommand{\chVII}[1]{\textcolor{BrickRed}{#1}}
\fi

\newcommand{\paratitle}[1]{\vspace{6pt}\textbf{#1.}}

\newcommand{\incircle}[1]{%
    \IfEqCase{#1}{%
        {1}{\ding{182}}%
        {2}{\ding{183}}%
        {3}{\ding{184}}%
        {4}{\ding{185}}%
        {5}{\ding{186}}%
        {6}{\ding{187}}%
        {7}{\ding{188}}%
        {8}{\ding{189}}%
    }[\PackageError{incircle}{Undefined option to incircle: #1}{}]%
}%

\crefformat{footnote}{#2\footnotemark[#1]#3}

\begin{document}

\begin{frontmatter}

\begin{fmbox}
\dochead{Methodology} 


\title{GRIM-Filter: Fast Seed Location Filtering \\in DNA Read Mapping \\Using Processing-in-Memory Technologies}


\author[
   addressref={cmu1,eth},
   email={jeremiekim123@gmail.com, calkan@cs.bilkent.edu.tr, onur.mutlu@inf.ethz.ch}
]{\inits{JSK}\fnm{Jeremie S.} \snm{Kim}}
\author[ 
   addressref={cmu1}, 
   email={dsenol@andrew.cmu.edu}
]{\inits{DS}\fnm{Damla} \snm{Senol Cali}}
\author[
   addressref={cmu2}, 
]{\inits{HX}\fnm{Hongyi} \snm{Xin}}
\author[
   addressref={nvidia}, 
]{\inits{DL}\fnm{Donghyuk} \snm{Lee}}
\author[
   addressref={cmu1}, 
]{\inits{SG}\fnm{Saugata} \snm{Ghose}}
\author[
   addressref={bilkent},
]{\inits{MA}\fnm{Mohammed} \snm{Alser}}
\author[
   addressref={eth}, 
]{\inits{HH}\fnm{Hasan} \snm{Hassan}}
\author[
   addressref={tobb}, 
]{\inits{OE}\fnm{Oguz} \snm{Ergin}}
\author[
   addressref={bilkent}, 
]{\inits{CA}\fnm{Can} \snm{Alkan*}}
\author[
   addressref={eth, cmu1}, 
   email={omutlu@gmail.com}
]{\inits{OM}\fnm{Onur} \snm{Mutlu*}}


\address[id=cmu1]{
  \orgname{Department of Electrical and Computer Engineering, Carnegie Mellon University}, 
  \city{Pittsburgh, PA},                              
  \cny{USA}                                    
}
\address[id=cmu2]{%
  \orgname{Department of Computer Science, Carnegie Mellon University}, 
  \city{Pittsburgh, PA}, 
  \cny{USA}
}
\address[id=nvidia]{%
  \orgname{NVIDIA Research},
  \city{Austin, TX},
  \cny{USA}
}
\address[id=bilkent]{%
  \orgname{Department of Computer Engineering, Bilkent University, Bilkent, Ankara, Turkey}, 
  \city{Bilkent}, 
  \cny{TR} 
}
\address[id=tobb]{%
  \orgname{Department of Computer Engineering, TOBB University of Economics and Technology}, 
  \city{Sogutozu}, 
  \cny{TR}
}
\address[id=eth]{%
  \orgname{Department of Computer Science, ETH Z{\"u}rich}, 
  \city{Z{\"u}rich}, 
  \cny{CH}
}


\begin{artnotes}
\end{artnotes}

\end{fmbox}


\begin{abstractbox}

\begin{abstract} 
\parttitle{Motivation} 
\chI{Seed location filtering} is critical in DNA read mapping, a process where billions of
DNA fragments (reads) sampled from a donor are mapped onto a reference genome
to identify genomic variants of the donor. State-of-the-art read mappers 1)
quickly generate possible mapping locations for seeds \chI{(i.e., smaller segments)} within each read, 2)
extract reference sequences at each of the mapping locations, and 3) check
similarity between each read and its associated reference sequences with a
\chI{computationally-expensive} algorithm (i.e., sequence alignment) to determine the
origin of the read. A \chI{seed location filter} comes into play before alignment,
discarding seed locations that alignment would deem a poor match. The
ideal \chI{seed location filter} would discard all poor match locations prior to
alignment such that there is no wasted computation on \chI{unnecessary} alignments.

\parttitle{Results} 
We propose a novel \chI{seed location filtering} algorithm, GRIM-Filter, optimized to exploit
\chII{3D-stacked memory systems} that integrate computation within
\chII{a logic layer stacked under memory layers}, to perform processing-in-memory (PIM).  GRIM-Filter quickly
filters seed locations by \chI{1)}~introducing a new representation of coarse-grained
segments of the reference genome, and \chI{2)}~using massively-parallel
in-memory operations to identify read presence within each coarse-grained
segment. Our evaluations show that for a sequence alignment error tolerance of
\chII{0.05}, GRIM-Filter \chII{1)~reduces the false negative rate of filtering by 5.59x--6.41x, and
2)~provides an end-to-end read mapper speedup of 1.81x--3.65x, compared to a
state-of-the-art read mapper employing the best previous seed location filtering algorithm}.

\parttitle{Availability}
The code \chIII{is} available online at:
\href{https://github.com/CMU-SAFARI/GRIM}{https://github.com/CMU-SAFARI/GRIM}
\end{abstract}




\end{abstractbox}
%

\end{frontmatter}




\section{Introduction} 


Our understanding of human genomes today is affected by \chI{the ability of modern
technology} to quickly and accurately determine an individual's entire genome. The
human genome is comprised of a sequence of approximately 3~billion bases that
are grouped into deoxyribonucleic acids (\emph{DNA}), but today's machines can 
identify DNA \chI{only} in short sequences (\chI{i.e.,} \emph{reads}). \chI{Determining} a
genome requires \chI{three} stages: 1)~cutting the genome into many short reads,
2)~identifying the DNA sequence of each read, and 3)~mapping each read against
the reference genome in order to analyze the variations in the sequenced
genome. In this paper, we focus on improving \chI{the third stage}, often referred to as
\emph{read mapping}, \chII{which is a major computational bottleneck of a
modern genome analysis pipeline}.  Read mapping is performed computationally by \emph{read
mappers} after each read has been identified. 

\begin{figure}[h]
    \centering
    \includegraphics[width=0.7\linewidth]{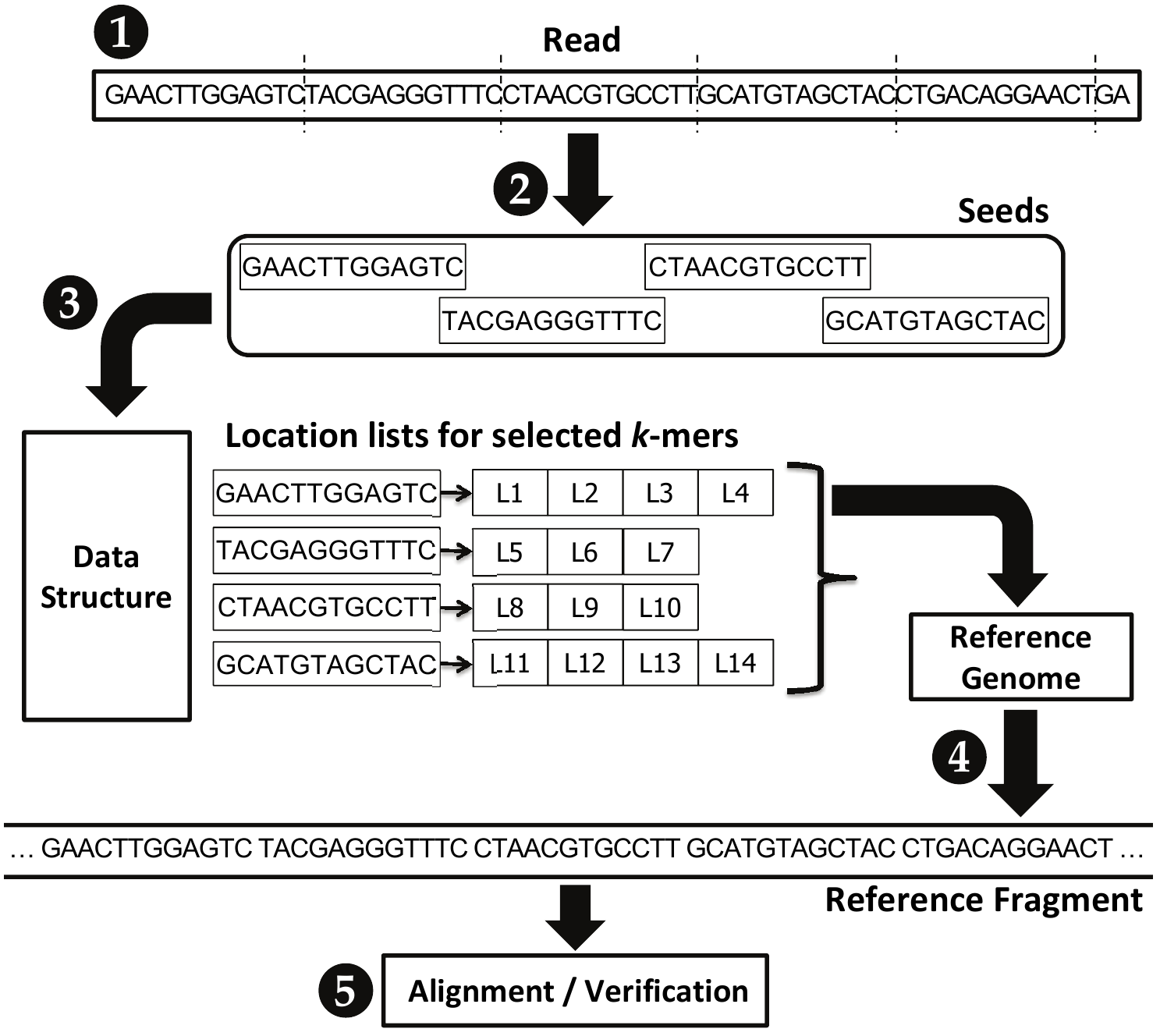}
    \caption{Flowchart of a seed-and-extend mapper.} 
    \label{fig:fastHASH}
\end{figure}

\chI{\emph{Seed-and-extend} mappers~\cite{hach2014mrsfast, ahmadi2012hobbes,
alkan2009personalized, rumble2009shrimp, hormozdiari2011sensitive,
weese2009razers} are a class of read mappers that break down
each read sequence into \emph{seeds} (i.e., smaller segments) to find locations in the 
reference genome that closely match the read.  Figure~\ref{fig:fastHASH} 
illustrates the five steps used by a seed-and-extend mapper.
First, the mapper obtains a read (\incircle{1} in the figure).
Second, the mapper selects smaller DNA segments from the read to serve as
seeds (\incircle{2}).
Third, the mapper indexes a data structure with \chII{each seed} to obtain a list of
possible locations within the reference genome that \chII{could} result in a match
 (\incircle{3}).
Fourth, for each possible location \chII{in the list}, the mapper obtains the
corresponding DNA sequence from the reference genome (\incircle{4}).  Fifth,
the mapper aligns the read sequence to the reference sequence (\incircle{5}),
using an expensive sequence alignment \chII{(i.e., verification)} algorithm to
determine the similarity between the \chIII{read sequence and the reference
sequence}.}


To improve the performance of seed-and-extend mappers, we can
utilize {\em seed location filters}, recently introduced by Xin et
al.~\cite{xin2013accelerating}. A seed location filter efficiently determines whether
a candidate mapping location would result in an incorrect mapping \emph{before}
performing the computationally-expensive sequence alignment step for that
location.  As long as the filter can eliminate possible locations \chII{that would result
in an incorrect mapping \emph{faster}} than
the time it takes to perform the alignment, the entire read mapping process can
be \chII{substantially accelerated~\cite{xin2015shifted, xin2013accelerating, alser2016gatekeeper, alser2017magnet}}.  As a result,
several recent works have focused on optimizing the performance of seed location 
filters~\cite{tran2015amas, xin2015shifted, xin2013accelerating,
xin2015optimal, alser2016gatekeeper, alser2017magnet}.


With the advent of seed location filters, the performance bottleneck of DNA
read mapping has shifted from sequence alignment to seed location
filtering~\cite{xin2013accelerating, xin2015shifted, alser2016gatekeeper,
alser2017magnet}. Unfortunately, a seed location filter requires large amounts
of memory bandwidth to process and characterize each of the candidate
locations. Our goal is to reduce the time spent in filtering and thereby
improve the speed of DNA read mapping. To this end, we present a new algorithm,
\chII{\emph{GRIM-Filter}}, to efficiently filter locations with high parallelism. We design
GRIM-Filter such that it is well-suited for implementation on 3D-stacked
memory, exploiting the parallel \chII{and low-latency} processing capability in the logic layer
\chI{of the memory}. 

3D-stacked DRAM\chII{~\cite{AMD-HBM, lee2015simultaneous, AMD-R9-Graphics,
o2014highlights, altera-HMC-UG, HMC-sources, hsieh2016transparent,
hsieh2016accelerating, ahn2015scalable, loh20083d}} is a \chII{new} technology
that integrates logic and memory in a \chII{three-dimensional} stack of dies with a large internal
data transfer bandwidth. This enables the bulk transfer of data from \chII{each memory layer} to
a logic layer that can perform simple parallel operations on the data. 

\chII{Conventional} computing requires the movement of data on \chII{the long, slow, and energy-hungry} buses between
\chII{the CPU processing cores} and memory such that cores can operate on data.  \chII{In contrast,} processing-in-memory
(PIM)-enabled devices such as 3D-stacked memory can perform simple arithmetic
operations very close to where the data resides, with high bandwidth and low
latency.  With carefully designed algorithms for PIM, application performance
can often be greatly improved \chII{(e.g., as shown in 
\chIII{\cite{ahn2015scalable, hsieh2016transparent, seshadri2017ambit, hsieh2016accelerating}})} 
\chII{because the relatively narrow and long-latency
bus between the CPU cores} and memory no longer impedes the speed of computation on the
data. 

\textbf{Our goal} is to develop a seed location filter that exploits the high memory
bandwidth and processing-in-memory capabilities of 3D-stacked DRAM to improve
the performance of DNA read mappers.

To our knowledge, this is the \textbf{first} seed location filtering algorithm
that accelerates read mapping by overcoming the memory bottleneck with PIM
using 3D-stacked memory technologies. GRIM-Filter can be used with any read
mapper.  However, in this work we demonstrate the effectiveness of GRIM-Filter
with a hash table based mapper, mrFAST with
FastHASH~\cite{xin2013accelerating}. \chI{We improve the performance of hash table
based read mappers while maintaining their high sensitivity and
comprehensiveness (which were \chII{originally demonstrated in} \cite{alkan2009personalized}).} 

\textbf{Key Mechanism.} GRIM-Filter provides a quick method for determining
whether a read will \textbf{not} match at a given location, thus allowing the
read mapper to skip the expensive sequence alignment process for that location.
GRIM-Filter works by counting the existence of small segments of a read in a
genome region. If the count falls under a threshold, indicating that many small
segments of a read are \chII{\emph{not}} present, GRIM-Filter discards the
locations in that region before alignment. The existence of all small segments
in a region are stored in a bitvector, which can be easily predetermined for
each region of a reference genome.  \chI{The bitvector for a reference genome
region is retrieved when a read must be checked for a match in the region.} We
find that this regional approximation technique not only enables a high
performance boost via high parallelism, but also improves filtering accuracy
over the state-of-the-art. \chIII{The filtering accuracy improvement comes from the
finer granularity GRIM-Filter uses in counting the subsequences of a read
in a region of a genome, compared to the state-of-the-art filter~\cite{xin2013accelerating}.}

\textbf{Key Results.} We evaluate GRIM-Filter qualitatively and quantitatively
against the state-of-the-art seed location filter,
\emph{FastHASH}~\cite{xin2013accelerating}. Our results show that GRIM-Filter
provides a \emph{5.59x--6.41x} smaller false negative rate (i.e., the
proportion of locations that pass the filter, but \chII{that} truly result in a
poor match during sequence alignment) than the best previous filter with
\emph{zero} false positives (i.e., the number of locations that do not pass the
filter, but \chII{that truly} result in a good match during sequence
alignment).  GRIM-Filter \chII{provides} an end-to-end performance improvement
of \emph{1.81x--3.65x} over a state-of-the-art DNA read mapper, \emph{mrFAST}
with \emph{FastHASH}, for a set of real genomic reads, when we use a sequence
alignment error tolerance of \chII{0.05}. We also note that as we increase the
\chI{sequence alignment error tolerance}, the performance improvement of our
filter over the state-of-the-art increases. This makes \chIV{GRIM-Filter} more
effective and relevant for future-generation error-prone sequencing
technologies, such as nanopore sequencing~\cite{david2016nanocall,
senol2017nanopore}.

%
%
 
\section{Motivation and Aim} 

Mapping reads against a reference genome enables the analysis of the variations
in the sequenced genome. As the throughput of \chI{read mapping} increases,
more large-scale \chII{genome} analyses become possible. The ability to deeply
characterize and analyze genomes at a large scale could change medicine from a
reactive to a preventative and further personalized practice. In order to
motivate our method for improving the performance of read mappers, we pinpoint
the performance bottlenecks of modern-day mappers on which we focus our
acceleration efforts.  We find that across our data set (see
Section~\ref{sec:exp_method}), a state-of-the-art read mapper, mrFAST with
FastHASH~\cite{xin2013accelerating}, on average, spends 15\% of its execution
time performing sequence alignment on locations that are found to be a match,
and 59\% \chVII{of its} execution time performing sequence alignment on
locations that are discarded because they are not found to be a match (i.e.,
\emph{false locations}).

Our \emph{goal} is to implement a \chI{\emph{seed location filter}} that
reduces the wasted computation time spent \chII{performing sequence alignment
on} such false locations.  \chII{To this end, a seed location filter would
quickly determine} if a location will \chIII{\emph{not}} match the read and, if
so, \chII{it would avoid the sequence alignment} altogether. The
\chIII{\emph{ideal}} seed location filter correctly finds all false locations
without \chI{increasing the time required to execute read mapping}.
We find that such an ideal seed location filter would improve the
\emph{average} performance of mrFAST \chII{(with FastHASH)}
by \emph{3.2x}.  This speedup is primarily due to the reduced number of false
location alignments. In contrast, most prior works~\cite{aluru2014review,
arram2013hardware, arram2013reconfigurable, ashley2010clinical,
chiang2006hardware, hasan2007hardware, houtgast2015fpga,
mcmahon2008accelerating, olson2012hardware, papadopoulos2013fpga,
waidyasooriya2014fpga, blom2011exact, liu2012soap3, luo2013soap3,
manavski2008cuda} gain their speedups by implementing \chIV{all or part} of the
read mapper in specialized hardware or GPUs, \chII{focusing mainly on the
acceleration of the sequence alignment process, \emph{not} the \emph{avoidance}
of sequence alignment}.  These works \chII{that accelerate sequence alignment
provide} orthogonal solutions, and could be implemented together with \chI{seed
location filters, \chII{including GRIM-Filter,}} for additional performance
improvement \chIII{(see Section~\ref{sec:related} for more detail)}.


\section{GRIM-Filter} 
\label{sec:grim_filter}
\label{sec:filter}

We now describe our proposal for a new seed location filter, GRIM-Filter.
\chI{At a high level, the key idea of GRIM-Filter is to store and utilize
metadata on \emph{short segments} of the genome, i.e., segments on the order of
several hundred base pairs long,} in order to quickly determine if a read will
\textbf{not} result in a match at that genome segment.

\subsection{Genome Metadata Representation}
\label{sec:filter:data}

\begin{figure}[h]
  \centering
  \includegraphics[width=0.95\linewidth]{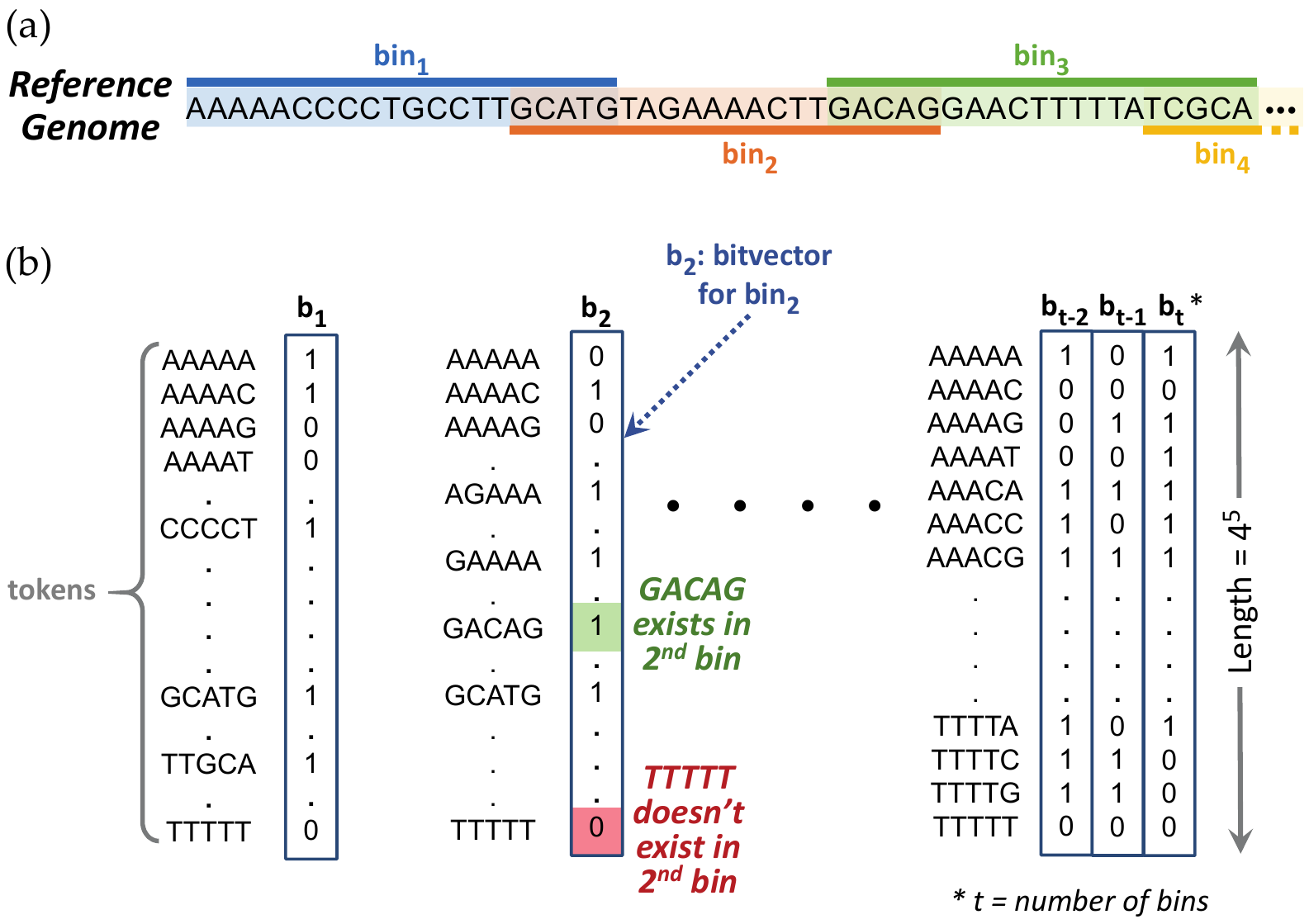}
  \caption{\chII{GRIM-Filter has a 2D data structure where each bit at $<$row, column$>$ indicates
  if a token (indexed by the row) is present in the corresponding bin (indicated by the column). 
  (a)~GRIM-Filter divides a genome into \chIII{overlapping} bins.
  (b)}~\chI{GRIM-Filter's metadata associated with a reference genome. Columns are indexed by the bin number of each location.} Rows are indexed by the \chII{token value}. In this figure, token size=$5$.}
  \label{fig:bitvectors}
\end{figure}

Figure~\ref{fig:bitvectors} shows a
reference genome with its associated metadata that is formatted for efficient
operation by GRIM-Filter. The reference genome is divided into short contiguous
segments, on the order of several hundreds of base pairs, which we refer to as
\emph{bins}.  \chI{GRIM-Filter operates at the granularity of these bins,
performing analyses on the metadata associated with each bin.  This metadata
is represented as a \emph{bitvector} that stores whether or not a \emph{token},
i.e., a short DNA sequence on the order of $5$ base pairs, is present within
the associated bin.} We refer to each bit \chIV{in the bitvector} as an \emph{existence bit}. To
account for all possible tokens of length $n$, each bitvector must be $4^n$
bits in length, where each bit denotes the existence of a particular token
instance.  Figure~\ref{fig:bitvectors} highlights the bits of two token
instances of $bin_2$'s bitvector\chI{: it shows that 1)~the token GACAG (green)
exists in $bin_2$, i.e., the \chIV{existence} bit associated with the token GACAG is set to 1 in
the $b_2$ bitvector; and 2)~the token TTTTT (red) is not present in $bin_2$,
i.e., the existence bit associated with the token TTTTT is set to 0 in the
$b_2$ bitvector.} 

\chI{Because these bitvectors are associated with the reference genome, the
bitvectors need to be generated only once per reference, and they can be used
to} map any number of reads from other individuals of the same species.
In order to generate the bitvectors, the genome must be \chI{sequentially
scanned for every possible token of length $n$, where $n$ is the selected token
size.  If $bin_x$ contains the token, the bit in the $b_x$ bitvector
corresponding to the token must be set ($1$). If $bin_x$ does not contain the
token, then the same bit is left unset (i.e., 0). These bitvectors are saved
and stored for later use when mapping reads to the same reference genome,
\chII{i.e., they are part of the genome's metadata}.}

\subsection{GRIM-Filter Operation}
\label{sec:filter:operation}

Before sequence alignment, 
\chI{GRIM-Filter checks each bin to see if the bin contains a potential mapping
location for the read, \chII{based on the \chIV{list of} potential locations provided by
the read mapper.  If the bin contains a location, GRIM-Filter then checks the
bin to see if \chIV{the location} is likely to match the read sequence}, by operating on the bitvector of the bin.}
\chI{This relies on the \emph{entire read} being contained within
a given bin, and thus requires the bins to overlap with each other
\chII{in the construction of the metadata} (i.e., some base pairs are
contained in multiple bins), as shown in \chII{Figure~\ref{fig:bitvectors}a}}. 

GRIM-Filter uses the described bitvectors to \chI{\emph{quickly}} determine whether
a match within a given error tolerance is impossible. This is done before
running the expensive sequence alignment algorithm, in order to reduce the
number of unnecessary sequence alignment operations. For each location \chII{associated with a seed}, 
\chI{GRIM-Filter 1)~loads the bitvector of the bin containing the location; 
2)~operates on the bitvector (as we will describe shortly) to quickly
determine if there will be no match \chII{(i.e., a poor match, given the error tolerance threshold)}; and 3)~discards the location if
it determines a poor match.} If GRIM-Filter does \textbf{not} discard the
location, the sequence at that location \textbf{must} be aligned with the read
to determine the match similarity. 

\begin{figure}[h]
  \centering
  \includegraphics[width=0.95\linewidth]{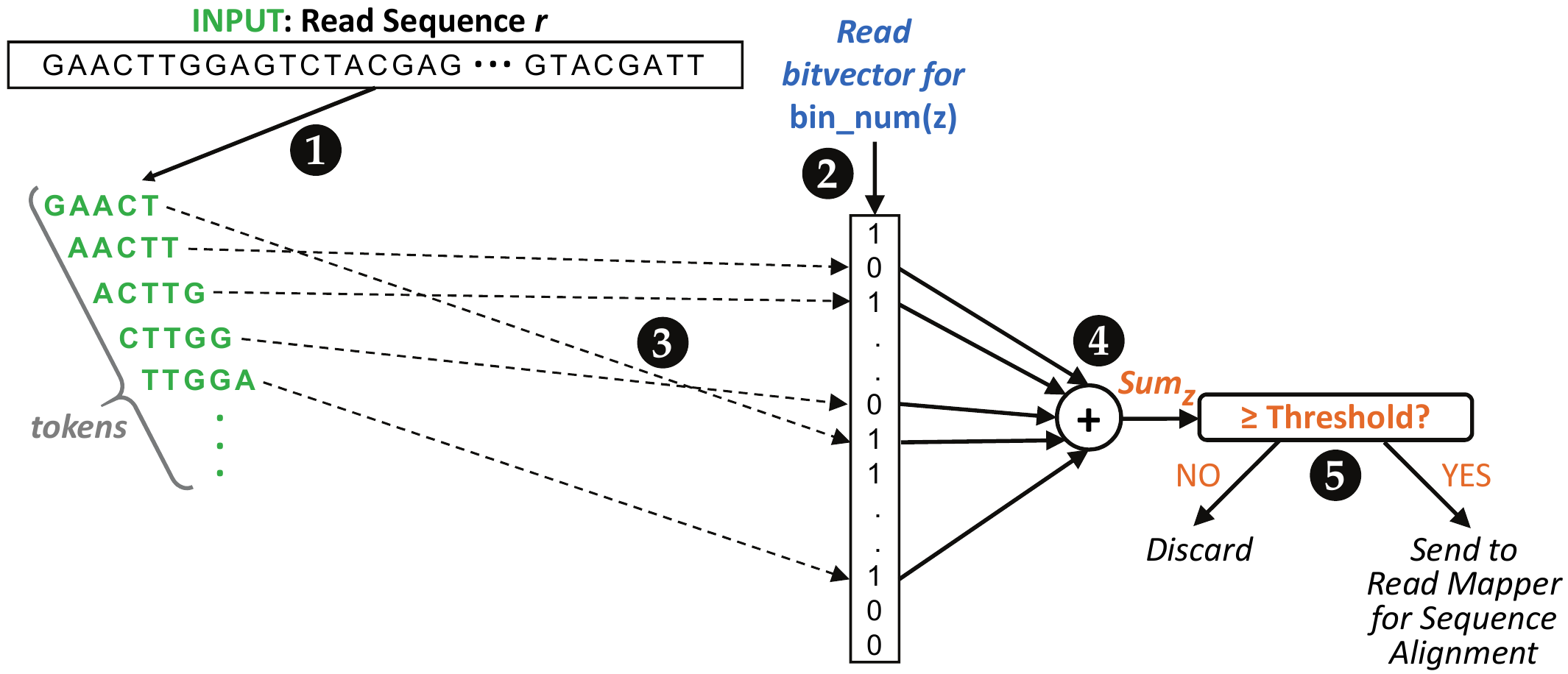}
  \caption{Flow diagram for our seed location filtering algorithm. GRIM-Filter takes in a read sequence and sums the existence of its tokens within a bin to determine whether \chII{1)}~the read sequence must be sequence aligned to the reference sequence in the bin \chII{or 2)~it can be discarded without alignment}.  \chV{Note that token size = 5 in this example.}} 
  \label{fig:GRIM_one_location}
\end{figure}

\chI{Using the circled steps in Figure~\ref{fig:GRIM_one_location}, we explain
in detail how GRIM-Filter determines whether to discard a location $z$ for a
read sequence $r$.
We use $bin\_num(z)$ to indicate the number of the bin that contains location $z$.
GRIM-Filter extracts every token contained within the read sequence $r$
(\incircle{1} in the figure).
Then, GRIM-Filter loads the bitvector for $bin_{bin\_num(z)}$ (\incircle{2}).
For each of the tokens contained in $r$, GRIM-Filter extracts the
existence bit of that token from the bitvector (\incircle{3}), to see whether
the token exists somewhere within the bin.
GRIM-Filter sums all of the extracted existence bits together (\incircle{4}),
which we refer to as the \emph{accumulation sum} for location $z$ ($Sum_z$).
The accumulation sum represents the number of tokens from read sequence $r$
that are present in $bin_{bin\_num(z)}$.  A larger accumulation sum indicates
that more tokens from $r$ are present in the bin, and therefore the location is
more likely to contain a match for $r$.
Finally, GRIM-Filter compares $Sum_z$ with a constant \emph{accumulation sum threshold}
value (\incircle{5}), to determine whether location $z$ is
likely to match read sequence $r$.  If $Sum_z$ is greater than or equal to the
threshold, then $z$ is likely to match $r$, and the read mapper must perform
sequence alignment on $r$ to the reference sequence at location $z$.  If
$Sum_z$ is less than the threshold, then $z$ \chVII{\emph{will not}} match $r$, and
the read mapper skips sequence alignment for the location.
We explain how we determine the accumulation sum threshold \chIII{in
Section~\ref{sec:filter:threshold}}.}


\chI{Once GRIM-Filter finishes checking each location, it returns control to the 
read mapper, which performs sequence alignment on \chII{\emph{only}} those locations that
pass the filter. This process is repeated for all seed locations, and \chIII{it}
significantly reduces the number of alignment operations, ultimately reducing
the end-to-end read mapping runtime \chII{(as we show in
Section~\ref{sec:results-analysis})}.} \chII{Our implementation of GRIM-Filter
ensures a zero \emph{false positive rate} (i.e., \chIII{no} locations that
result in correct mappings for the read sequence are incorrectly rejected by
the filter), \chIV{as GRIM-Filter passes any seed location whose bin contains
\chVII{enough of} the same tokens as the read sequence}.
GRIM-Filter can also account for errors in the sequence, \chV{when some of the
tokens do not match perfectly} (see \chIII{Section~\ref{sec:filter:errors}}).
Therefore, using GRIM-Filter to filter out seed locations does \emph{not}
affect the correctness of the read mapper.}

\subsection{Integration with a Full Read Mapper}
\label{sec:filter:integration}

\begin{figure}[h]
  \centering
  \includegraphics[width=0.95\linewidth]{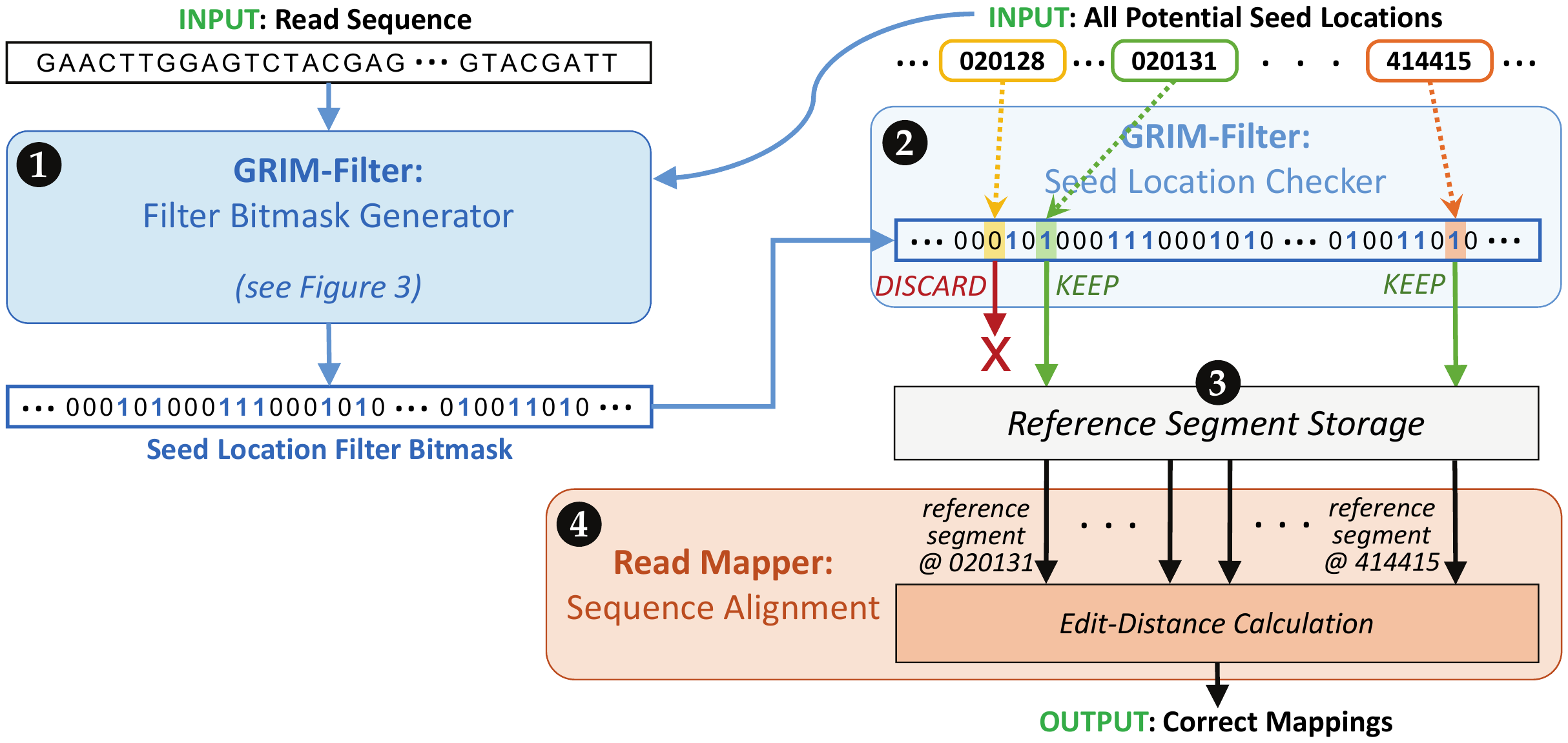}
  \caption{\chII{GRIM-Filter integration with a read mapper.  The \chIII{\emph{Filter Bitmask Generator}} uses
  the bitvectors for each bin to determine whether any locations within the bin
  are potential matches with the read sequence, and saves potential match 
  information into a \chV{\emph{Seed Location Filter Bitmask}}.  The \chIII{\emph{Seed Location Checker}} uses the bitmask to
  retrieve the corresponding reference segments for only those seed locations that match,
  which are then sent to the read mapper for sequence alignment.}} 
  \label{fig:GRIM_integration} 
\end{figure}

\chII{Figure~\ref{fig:GRIM_integration} shows how we integrate GRIM-Filter with
a read mapper to improve read mapping performance.  Before the read mapper
begins sequence alignment, it sends the read sequence, along with all potential
seed locations found in the hash table for the sequence, to GRIM-Filter.  Then,
the \chIII{\emph{Filter Bitmask Generator}} for GRIM-Filter performs the seed
location filtering algorithm we describe \chIII{in
Section~\ref{sec:filter:operation}}, checking only the bins that include a
potential seed location to see if the bin contains the same tokens as the read
sequence (\incircle{1} in \chV{Figure~\ref{fig:GRIM_integration}}).  For each location, we save the output
of our threshold decision \chIII{(the computation of which was shown in
Figure~\ref{fig:GRIM_one_location})} as a bit within a \emph{seed location
filter bitmask}, where a 1 means that the location's accumulation sum was
greater than or equal to the threshold, and a 0 means that the accumulation sum
was less than the threshold.  This bitmask is then passed to the
\chIII{\emph{Seed Location Checker}} (\incircle{2} in \chV{Figure~\ref{fig:GRIM_integration}}), which locates the
reference segment corresponding to each seed location that passed the filter
(\incircle{3}) and sends the reference segment to the read mapper.  The read
mapper then performs sequence alignment on \chIII{\emph{only}} the reference
segments it receives from the seed location checker (\incircle{4}), and outputs
the correct mappings for the read sequence.}

\subsection{Determining the Accumulation Sum Threshold}
\label{sec:filter:threshold}

\chI{We now discuss in detail how to determine 
the threshold used to evaluate the accumulation sum ($Sum_z$). 
\chIII{The threshold is used to determine whether or not a seed location should
be sent to the read mapper for sequence alignment (\chV{shown as
\incircle{5} in Figure~\ref{fig:GRIM_one_location}})}.  A greater value of
$Sum_z$ indicates that \chIV{the seed location $z$} is more likely to be a good match
for the read sequence $r$.  However, there are cases where $Sum_z$ is high, but
the read sequence results in a poor match \chIV{with the seed location $z$}. A
simple example of this poor match is a read sequence that consists entirely
of ``A'' base pairs, resulting in 100 AAAAA tokens, and a \chIV{seed location}
 that consists entirely of ``G'' base pairs except for a single
AAAAA token.  In this example, all 100 AAAAA tokens in the read sequence
locate the one AAAAA token in the \chIV{seed location}, resulting in an accumulation sum of
100, even though the \chIV{location} contains only one AAAAA token.  Because such cases
occur, even though they \chII{may} occur with low probability, GRIM-Filter
cannot \textbf{guarantee} that a high accumulation sum for a \chIV{seed location}
corresponds to a good match with a read sequence.  On the other hand,
GRIM-Filter can guarantee that a low accumulation sum (i.e., a sum that
falls under the threshold) indicates that any reference sequence
within the bin is a poor match with the read sequence.
This is because a lower sum means that fewer tokens from the read sequence are
present in the bin, which translates directly to a greater number of errors in a
potential match.  For a low enough sum, we can guarantee that the potential
read sequence alignment would have too many errors to be a good match.

\subsection{Taking Errors into Account}
\label{sec:filter:errors}

\chI{If a read maps perfectly to a reference sequence in $bin_{bin\_num(z)}$,
$Sum_z$ would simply be the total number of tokens in a read, which is
$read\_length~-~(n-1)$ for a token size of $n$. 
However, to account for insertions, deletions, and substitutions in the \chIV{read} sequence,
sequence alignment has some error tolerance, where a read sequence and a
reference sequence are considered a good match \chIII{\emph{even if}} some differences exist.
The accumulation sum threshold must account for this error tolerance, so
we reduce the threshold below $read\_length~-~(n-1)$ to allow some tokens
to include errors.
Figure~\ref{fig:Threshold_eq}a shows the equation that we use to calculate the
threshold while accounting \chII{for errors}.
As shown in \chIII{Figure~\ref{fig:Threshold_eq}b}, a token of size $n$
in a bin overlaps with $n-1$ other tokens. We calculate the lowest $Sum_z$
possible for a sequence alignment that includes only a
single error (i.e., one insertion, deletion, or substitution) by
studying these $n$ tokens.  
If the error is an insertion, the insertion shifts at least one of the $n$
tokens to the right, preserving the shifted token while changing the
remaining tokens ($n-1$ in the worst case).
If the error is a deletion or a substitution, the change in the worst case can
affect all $n$ tokens.
\chIII{Figure~\ref{fig:Threshold_eq}b shows an
example of how a substitution affects four different tokens, where $n = 4$.}
Therefore, for each error that we tolerate, we must assume \chII{the worst-case
error (i.e., a deletion or a substitution), in which case} up to
$n$ tokens will \chII{\emph{not} match with the read sequence even when
the location actually contains the read sequence}.}

\begin{figure}[h]
  \centering
  \includegraphics[width=0.9\linewidth]{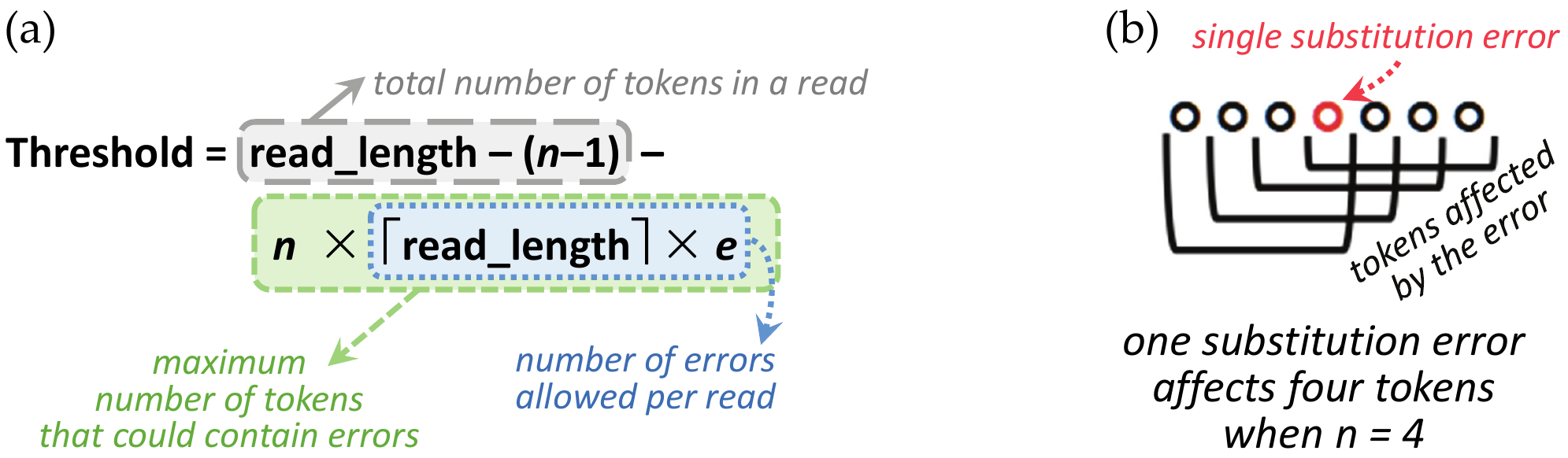}
  \caption{\chII{(a)}~Equation to \chI{calculate the accumulation sum threshold for a read sequence, 
where $n$ is the token length and $e$ is the sequence alignment error tolerance.} 
\chII{(b)}~\chIII{Impact of a substitution error on four separate tokens, when $n=4$.}  A single deletion or substitution error propagates to 4~consecutive tokens, while a single insertion error propagates to 3~consecutive tokens.}
  \label{fig:Threshold_eq}
\end{figure}

\chI{The equation in Figure~\ref{fig:Threshold_eq}a gives the \chIII{accumulation sum} threshold,
accounting for the worst-case scenario for a sequence alignment error tolerance of $e$.
This means that the maximum number of allowable errors is equal to the \chII{\emph{ceiling}} of the
read size multiplied by the sequence alignment error tolerance.
A sequence alignment error tolerance of \chII{$e=0.05$} or less is
widely used~\cite{ahmadi2012hobbes, cheng2015bitmapper, hatem2013benchmarking,
xin2015shifted}.  For each allowable error, we assume that the worst-case number
of tokens (equal to the token length $n$) are affected by the error.
We also assume the worst case that \chII{each error affects a different set of
tokens within the \chIII{read},}
which results in the greatest possible number of tokens that may not match.
We calculate this by multiplying the maximum number of allowable errors by $n$
in the equation.
Finally, we subtract the \chIII{largest} possible number of tokens that may \chIII{\emph{not}} match
from the total number of tokens in the read sequence, which is $read\_length~-~(n-1)$.}
\chIII{This leads to the threshold value that GRIM-Filter uses to determine the seed locations that
the read mapper should perform sequence alignment on, as discussed in
Section~\ref{sec:filter:operation} and \chV{shown as
\incircle{5} in Figure~\ref{fig:GRIM_one_location}}.}

\subsection{Candidacy for 3D-Stacked Memory Implementations}
\label{sec:filter:3d}

\chI{We identify \textbf{three} characteristics of \chII{the filter bitmask generator in} GRIM-Filter that make it a
\chII{strong} candidate for implementation in 3D-stacked memory: 1)~it requires only
very simple operations (\chII{e.g., sums and comparisons)}; 2)~it is highly
parallelizable, since each bin can be operated on independently and in parallel;
and 3)~it is highly memory-bound, requiring a single memory access for
approximately every three computational instructions \chII{(we determine this
by profiling a software implementation of GRIM-Filter, \chIII{i.e., GRIM-Software,
which is described in Section~\ref{sec:exp_method}})}.}
\chI{Next, we describe how we implement GRIM-Filter in 3D-stacked memory.}


\section{Mapping GRIM-Filter to 3D-Stacked Memory}

\chI{In this section, we first describe \chII{the} \emph{3D-stacked DRAM} technology \chIII{(Section~\ref{subsec:3d_memory})}, which attempts to bridge the \chII{well-known}
disparity between processor speed and memory bandwidth. Next, we describe how
GRIM-Filter can be easily mapped to utilize this new memory technology
(Section~\ref{subsec:map_grim_filter}). As the disparity between processor
speed and memory bandwidth increases, memory becomes more of a bottleneck in
the computing stack in terms of both performance and energy
consumption~\cite{mutlu2003runahead, mutlu2014research, mutlu2013memory, ipek2008self, ahn2015scalable}.}
Along with 3D-stacked DRAM, which enables much higher bandwidth and lower
latency compared to conventional DRAM, the disparity between processor and
memory is alleviated by the re-emergence of the concept of
\emph{Processing-in-Memory} \chI{(PIM).  PIM} integrates processing units inside or near
the \chII{main memory} to \chIII{1)}~leverage high \chIII{in/near-DRAM} bandwidth, and \chIII{low intra-DRAM latency}; and \chIII{2)}~reduce energy
consumption by reducing the amount of data transferred to \chII{and from} the processor.  In
this section, we briefly explain the required background for these two
technologies, \chI{which we leverage to implement GRIM-Filter in a \chIII{highly-parallel} manner.}


\subsection{3D-Stacked Memory}
\label{subsec:3d_memory} 


\chI{Main memory is implemented using the DRAM (dynamic random access memory)
technology \chII{in} today's systems\chII{~\cite{kim2015ramulator, mutlu2015main, kim.book14}}.
Conventional DRAM chips are connected to the processors using \chIII{long, slow, and energy-hungry PCB (printed circuit board)}
interconnects\chII{~\cite{kim2012case, raidr, lee2013tiered, kim2015ramulator,
seshadri2017simple, kim.book14, lee2015decoupled}}.  \chIV{The conventional DRAM chips} do not incorporate logic to perform computation. \chII{For more detail on \chIII{modern} DRAM operation and architecture, we refer the reader to our previous works \chIV{(e.g., \cite{hassan2016chargecache, lee2017design, lee2015adaptive, lee2013tiered, kim2014flipping, kim2012case, chang2017understanding, chang2016understanding, raidr, liu2013experimental, kim2015ramulator, lee2015decoupled, hassan2017softmc, patel2017reach, chang2014improving, chang.thesis17, kim.thesis15, lee.thesis16})}.}

3D-stacked DRAM is a new DRAM technology that has a much higher internal
bandwidth than conventional DRAM, thanks to the closer integration of logic and
memory using the \emph{through-silicon via} (TSV) interconnects,} as seen in
Figure~\ref{fig:HBM}.  TSVs are \chIII{new,} vertical interconnects that can pass through
the silicon wafers of a 3D stack of dies\chII{~\cite{kim2009study, loh20083d, lee2015simultaneous}}. A TSV has a much
smaller feature size than a \chIII{traditional} PCB interconnect, which enables a 3D-stacked DRAM
to integrate hundreds to thousands of these wired connections between stacked
layers. Using \chI{this large number of} wired connections, 3D-stacked DRAM can transfer bulk
data simultaneously, enabling much higher bandwidth compared to conventional
DRAM.  Figure~\ref{fig:HBM} shows a 3D-stacked DRAM (e.g., High Bandwidth
Memory \cite{AMD-HBM, standard2013high}) based system that consists of
\chII{four layers of DRAM dies and a logic die stacked together and connected} using TSVs, a processor die, and a silicon interposer that
connects the stacked DRAM and the processor. The vertical connections in the
stacked DRAM are very wide and very short, which results in {\em high bandwidth}
and {\em low power consumption}, respectively~\cite{lee2015simultaneous}.
\chI{There are many different 3D-stacked DRAM architectures available today.
High Bandwidth Memory (HBM) is already integrated into the AMD Radeon R9
Series graphics cards~\cite{AMD-R9-Graphics}. \chII{High} Bandwidth Memory~2
(HBM2) is integrated in \chII{both} the new AMD Radeon RX Vega$^{64}$ Series
graphics cards~\cite{AMD-RX-Vega-Graphics} and \chII{the} new NVIDIA Tesla P100 GPU
accelerators\chII{~\cite{NVIDIA-P100}}.  Hybrid Memory Cube (HMC) is developed by a number of different
contributing companies~\cite{altera-HMC-UG, HMC-sources}. Like HBM, HMC also
enables a logic layer underneath the DRAM layers that can perform 
\chII{computation~\cite{ahn2015scalable, hsieh2016transparent, hsieh2016accelerating}}. HMC is already integrated in the
SPARC64 XIfx chip~\cite{yoshida2014sparc64}.   Other new technologies
that can enable processing-in-memory are also already prototyped in real chips,
such as Micron's Automata Processor~\cite{dlugosch2014efficient} and Tibco
transactional application servers~\cite{tibco-imc, micron-automata}.} 

\begin{figure}[h]
    \centering
    \includegraphics[width=0.95\linewidth]{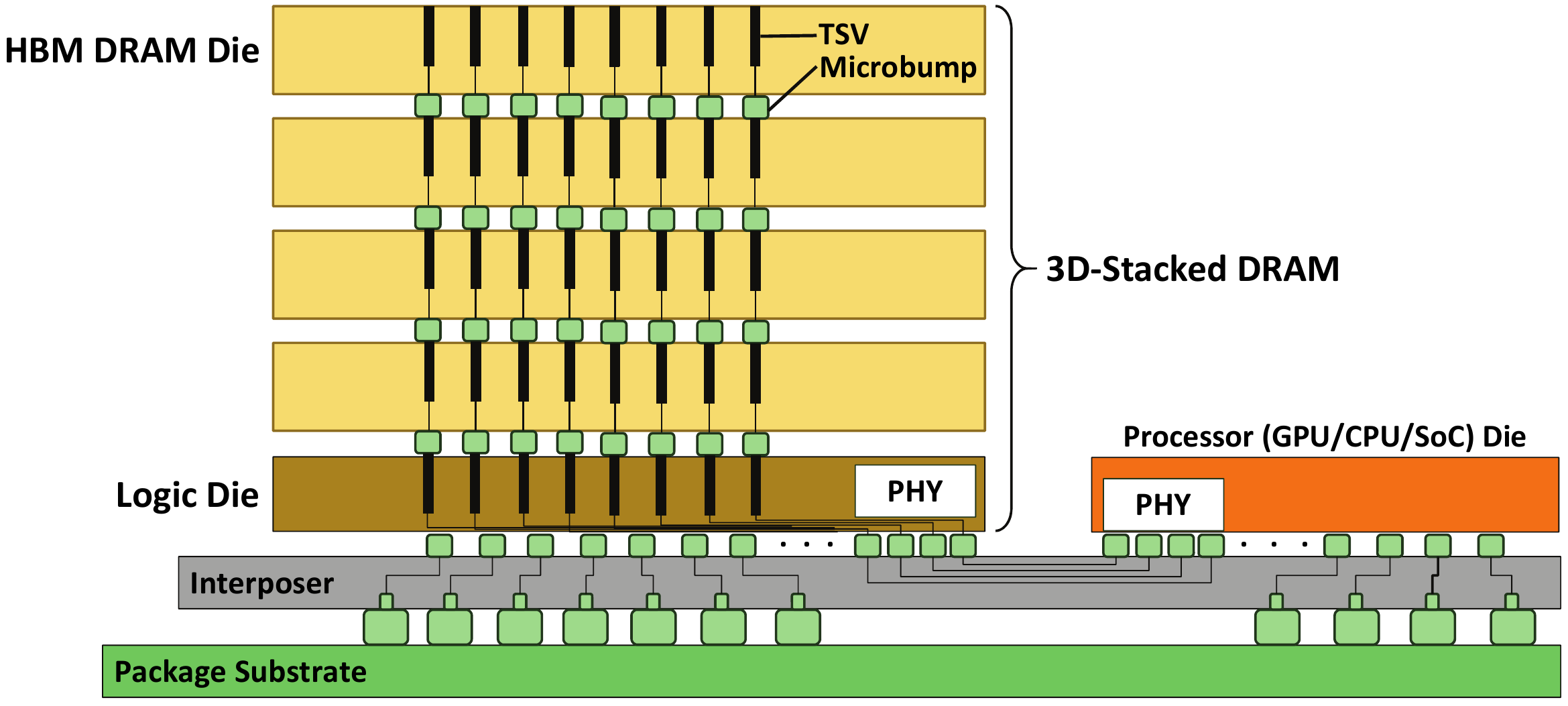}
    \caption{3D-stacked DRAM example. High Bandwidth Memory consists of stacked memory layers \chII{(four layers in the picture)} and a logic layer connected by high bandwidth \chI{through-silicon vias (TSVs)} and microbumps\chII{~\cite{AMD-HBM, standard2013high, lee2015simultaneous}}. The 3D-stacked memory is then connected to \chI{a processor die} with an interposer layer that provides high-bandwidth between the logic layer and the processing units on the package substrate.}
    \label{fig:HBM}
\end{figure}

\paratitle{Processing-in-Memory (PIM)}
\chI{A key technique to improve
performance (both bandwidth and latency) and reduce energy consumption in the
memory system is to place computation units inside the memory system, where the
data resides. Today, we see processing capabilities appearing inside and near
DRAM memory (e.g., in the logic layer of 3D-stacked memory)~\cite{ahn2015scalable, ahn2015pim, lee2015simultaneous, seshadrifast,
seshadri2013rowclone, seshadri2015gather, seshadri2017ambit, liu2017concurrent,
seshadri2017simple, pattnaik2016scheduling, babarinsa2015jafar,
farmahini2015nda, gao2015practical, gao2016hrl, hassan2015near,
hsieh2016transparent, morad2015gp, sura2015data, zhang2014top,
hsieh2016accelerating, boroumand2017lazypim, chang2016low}. This \chIV{computation inside
or near DRAM} significantly reduces the need to transfer data \chII{to/from} the processor
over the memory bus.
PIM provides significant performance improvement and energy reduction compared
to the conventional system architecture\chIII{~\cite{ahn2015scalable, ahn2015pim, akin2015data, guo20143d, hsieh2016accelerating, seshadri2017ambit, hsieh2016transparent, boroumand2017lazypim}}}, which must transfer \emph{all data} \chII{to/from} the
processor since the processor \chII{is the only entity that} performs all computational tasks.

\paratitle{3D-Stacked DRAM with PIM} 
\chI{The combination of the two new
technologies, 3D-stacked DRAM and PIM, enables very promising opportunities to
build very high-performance and \chII{low-power} systems. A promising design for
3D-stacked DRAM consists of multiple stacked memory layers and a
tightly-integrated logic layer that controls the stacked memory, as shown in
Figure~\ref{fig:HBM}. As many prior works show\chII{~\cite{ahn2015scalable,
ahn2015pim, akin2015data, guo20143d, loh20083d, zhu20133d, zhu2013accelerating,
hsieh2016accelerating, lee2015simultaneous, hsieh2016transparent,
liu2017concurrent, boroumand2017lazypim}}, the logic layer in 3D-stacked DRAM can be utilized not only
for managing the stacked memory layers, but also for integrating
application-specific accelerators or simple processing cores. Since the logic
layer already exists and has enough space to integrate computation units,
integrating application-specific accelerators in the logic layer requires
modest design and implementation overhead, and little to no hardware overhead
\chII{(see \cite{zhang2014top, hsieh2016accelerating} for various analyses)}.
Importantly, the 3D-stacked DRAM architecture enables us to fully customize the
logic layer for the acceleration of applications using processing-in-memory
(i.e., processing in the logic layer)\chII{~\cite{ahn2015scalable, zhu20133d,
ahn2015pim, hsieh2016accelerating}}.} 

\subsection{Mapping GRIM-Filter to 3D-Stacked Memory with PIM} 
\label{subsec:map_grim_filter}

\chI{We find that GRIM-Filter is a \chII{very} good candidate to implement using
processing-in-memory, as the filter is memory-intensive and
performs simple computational operations (\chII{e.g., simple} comparisons and addition\chIII{s}). 
Figure~\ref{fig:3D_and_both} shows how we implement GRIM-Filter in a
3D-stacked memory.  The \emph{center block} shows each layer of an
example 3D-stacked memory \chIII{architecture}, where multiple DRAM layers are stacked above
a logic layer.  The layers are connected together with several hundred TSVs,
which enable a high data transfer bandwidth between the layers.
Each DRAM layer is subdivided into multiple \emph{banks} of memory.
A bank in one DRAM layer is connected to banks in the other DRAM layers using the TSVs. These interconnected banks, along with a slice of the logic layer,
are grouped together into a \emph{vault}.
Inside the 3D-stacked memory, we store the \chII{bitvector of each bin
(see Section~\ref{sec:grim_filter}) within a bank} \chIII{as follows: 1)~each bit
of the bitvector is placed in a different row in a consecutive manner
\chV{(e.g., bit~0 is placed in row~0, bit~1 in row~1, and so on)}; and
2)~all bits of the bitvector are placed in the same column, and the entire bitvector fits in the column
\chVI{(e.g., bitvector~0 is placed in column~0, bitvector~1 in column~1, and so on)}}.
We \chIII{design and} place customized logic to perform the GRIM-Filter operations within
each logic layer slice, such that each vault can perform \chIII{independent} GRIM-Filter
operations \chIV{in parallel with \chV{every other vault}}.  Next, we discuss how we organize the bitvectors
within each bank.  Afterwards, we discuss the customized logic required
for GRIM-Filter \chII{and the associated hardware cost}.}


\begin{figure*}[h]
  \centering
  \includegraphics[width=0.95\linewidth]{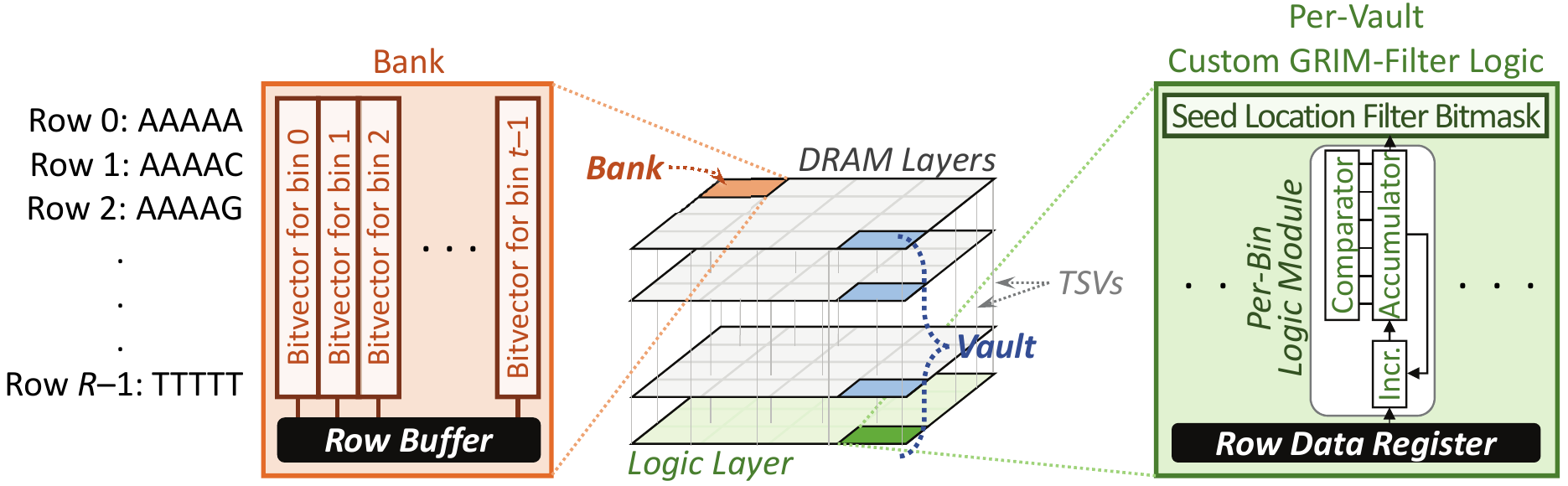} 
  \caption{\textit{Left block:} \chIII{GRIM-Filter bitvector layout within a DRAM bank. \textit{Center block:} 3D-stacked DRAM with tightly integrated logic layer stacked underneath with TSVs for a high intra-DRAM data transfer bandwidth. \textit{Right block:} Custom GRIM-Filter logic placed in the logic layer, \chV{for each vault}.}} 
  \label{fig:3D_and_both} 
\end{figure*}

\chI{
The \emph{left block} in Figure~\ref{fig:3D_and_both} shows the layout of
bitvectors in a single bank.  The bitvectors are written in column order
\chIII{(i.e., column-major order)} to the banks, such that \chIII{a}
DRAM access \chIII{to a row fetches the} existence bits of the same \chII{token
across \emph{many} bitvectors (e.g., bitvectors 0 to $t-1$ in the example in
Figure~\ref{fig:3D_and_both})}. \chIII{When GRIM-Filter reads a row of data 
from a bank, the DRAM buffers the row within the bank's \emph{row buffer}~\cite{kim2012case, lee2013tiered, mutlu.isca08,
mutlu.micro07},
which resides in the same DRAM layer as the bank.  
This data is then copied into a \emph{row data register} that sits in the logic layer,
from which the GRIM-Filter logic can read the data.}
This data organization allows
each vault to compute the accumulation sum of multiple bins \chII{(e.g., bins 0
to $t-1$ in the example)} simultaneously. \chII{Thus, GRIM-Filter can} quickly
and efficiently determine, across many
bins,\chII{\footnote{\label{ft:eq1}\chII{In other words, the number of bins 
\chV{that can be accessed in parallel from each bank,} times the number of banks times the number of vaults, within a DRAM
chip.}}} whether a \chIV{seed location} needs to be \chIII{discarded or} sequence aligned in
any of these bins.}

\chI{The \emph{right block} in Figure~\ref{fig:3D_and_both} shows the custom
hardware logic implemented for GRIM-Filter in each vault's logic layer.  We
design a small \chII{\emph{logic module}} for GRIM-Filter, which consists of
only an incrementer, accumulator, and comparator, and operates on the bitvector$_x$ 
of a single bin~$x$.  The incrementer adds 1 to the value in the accumulator, which
stores the accumulated sum for bin~$x$.  In order to hold the final sum \chV{(i.e., $Sum_z$, 
shown as \incircle{4} in Figure~\ref{fig:GRIM_one_location})}, each
accumulator must be at least $\lceil\log_2(read\_length)\rceil$ bits wide.
Each comparator must be \chII{of} the same width as the accumulator, as the
comparator is used to check whether the accumulated sum exceeds the accumulated
sum threshold.  Because of the way we arrange the bitvectors in DRAM, a single
read operation in a vault retrieves \chII{many (e.g., $t$)} existence bits in
parallel, \chIII{from \chII{many (e.g., $t$)} bitvectors, for the same token.  These
existence bits are copied from a DRAM bank's row buffer into a \emph{row data
register}} within the logic layer slice of the vault.  In order to
\chII{maximize throughput,} we add a GRIM-Filter \chIII{logic module \emph{for
each bin} to the logic layer slice.  This allows GRIM-Filter to process all of
these existence bits\cref{ft:eq1} from multiple bitvectors in parallel.}}

\paratitle{Integration into the System and Low-Level Operation}
\chI{When
GRIM-Filter starts in the CPU \chII{(spawned by a read mapper)}, it sends a
read sequence $r$ to the in-memory GRIM-Filter logic, along with a
\chII{\emph{range}} of consecutive bins to check for a match.
\chIII{GRIM-Filter quickly checks the range of bins to determine whether or not
to discard seed locations within those bins.} \chIV{In the logic layer, the
GRIM-Filter Filter Bitmask Generator (see Section~\ref{sec:filter:integration})}
iterates through each token in read sequence~$r$.  For each token,
GRIM-Filter reads the memory row in each vault that contains the existence bits
for that token, for the bins being checked, \chIV{into the row buffer inside
the DRAM layer. Then, GRIM-Filter copies the row to the row data register} in the \chII{logic
layer}.  \chII{Each} GRIM-Filter \chII{logic module is assigned to a single
bin.  The logic module examines the bin's existence bit in the row buffer},
and the incrementer adds one to the value in the accumulator only if the
existence bit is set.  This process is repeated for all tokens in $r$.
Once all of the tokens are processed, each \chII{logic module} uses its
comparator to check if the accumulator, which now holds the accumulated sum
($Sum_z$, \chV{shown as \incircle{4} in Figure~\ref{fig:GRIM_one_location}}) for \chII{its assigned bin}, is greater than or equal to the
accumulated sum threshold.  If $Sum_z$ is greater than or equal to the
threshold, a \emph{seed location filter bit} is set, indicating that the
read sequence should be sequence aligned with the locations in the bin by
the read mapper.  To maintain \chII{the same amount of parallelism present
\chIII{in} the bitvector operations}, we place the seed location filter bits into a
\chIII{\emph{seed location filter bitmask}}, where each \chII{logic module}
writes to one bit in the \chIII{bitmask} once it performs the accumulator sum
threshold comparison.  \chIII{The seed location filter bitmask is then
written to the DRAM layer.}  Once the \chIII{Seed Location Checker (see Section~\ref{sec:filter:integration}) 
starts executing in the CPU}, it reads
the \chIII{seed location filter bitmasks} from DRAM, and performs sequence
alignment for only those bits whose seed location filter bits are set to 1.

\paratitle{Hardware Overhead}
\chI{\chII{The hardware overhead \chIII{of our GRIM-Filter implementation in 3D-stacked
memory} depends on the available bandwidth $b$ between a memory layer and the
logic layer. In HBM2~\cite{o2014highlights}, this bandwidth is 4096~bits per
cycle \chIII{across all vaults}} (i.e., each clock cycle, 4096~bits \chII{from a memory layer can be
copied to the \chIII{row data registers}} in the logic layer). GRIM-Filter exploits all of
this parallelism \chII{completely}, as we can place \chII{$b$~GRIM-Filter logic
modules (4096~modules for HBM2)} \chIII{across all vaults within the logic
layer}.  In total, for an HBM2 memory, and for a read mapper that processes
reads consisting of 100~base pairs, GRIM-Filter requires 4096~incrementer
lookup tables (LUTs), 4096~seven-bit counters \chII{(a seven-bit counter can
hold the maximum accumulator sum for a 100-base-pair read sequence)},
4096~comparators, and \chII{enough} buffer space to hold the seed location
filter \chIII{bitmasks}.  \chV{With a larger bandwidth between the logic and memory
layers, we would be able to compute the \chIII{seed location filter bits} for
more bins in parallel, but this would also incur a larger hardware overhead in
the logic layer.}

\chIII{While the read mapper performs sequence
alignment on seed locations specified by one seed location filter bitmask,
GRIM-Filter generates seed location filter bitmasks for a different set of seed
locations.  We find that a bitmask buffer size of 512~KB \chV{(stored in DRAM)} provides enough
capacity to ensure that GRIM-Filter and the read mapper never stall due to a
lack of buffer space.} 

\chII{The \chIII{overall memory footprint (i.e., \chIV{the} amount of storage space
required)} of the bitvectors for a reference genome is calculated by
multiplying the number of bins by the size of a single bin.  In
Section~\ref{subsec:motivational-data}, we show how we find a set of parameters
that results in an effective filter with a low memory footprint (3.8~GB).}

\chII{We conclude that} GRIM-Filter requires a modest and simple logic layer,
which gives it an advantage over other \chIV{seed location filtering} algorithms that could be
implemented \chIV{in} the logic layer.}

\section{Experimental Methodology}
\label{sec:exp_method}

{\bf Evaluated Read Mappers.}
We evaluate our proposal \chI{by \chII{incorporating GRIM-Filter into}} the
state-of-the-art hash table based read mapper, mrFAST with
FastHASH~\cite{xin2013accelerating}. We choose this mapper for our evaluations
as it provides high accuracy in the presence of relatively many errors, which is
required to detect genomic variants within and across
species~\cite{alkan2009personalized, xin2013accelerating}. 
\chI{GRIM-Filter plugs in as an extension to mrFAST, using a simple series of
calls to an application programming interface (API).}
However, we note
that GRIM-Filter can be used with any other \chII{read} mapper. 

\chI{We evaluate two read mappers:}
\begin{itemize}
\item \chIV{\emph{mrFAST with FastHASH}~\cite{xin2013accelerating}, which does \emph{not} use GRIM-Filter;}


\item \chI{\emph{GRIM-3D}, our 3D-stacked memory implementation of GRIM-Filter 
combined with mrFAST and the non-filtering portions of FastHASH.}
\end{itemize}

\paratitle{Major Evaluation Metrics}
We report \chI{{1)}~GRIM-Filter's \emph{false
negative rate} (i.e., the fraction of locations \chII{that pass} through the filter \chII{but} do not
contain a match with the read sequence), and
{2)} the end-to-end \emph{performance improvement}} of the read mapper when using
GRIM-Filter. 
We measure the false negative rate of our filter (and the baseline filter used
by the mapper) as the ratio of the number of locations that passed the filter
but did not result in a mapping over all locations that passed the filter.
\chII{Note that our implementation of GRIM-Filter ensures} a zero \emph{false
positive rate} (i.e., \chIII{it does \emph{not} filter out any correct mappings
for the read sequence}), \chIII{and, thus,} GRIM-Filter does not \chII{affect} the
correctness of a read mapper. 

\paratitle{Performance Evaluation} 
\chI{We measure the performance improvement of GRIM-3D by comparing
the execution time of our read mappers.  
\chII{We \chIII{develop a methodology to} \emph{estimate} the performance of
GRIM-3D, since real hardware systems that enable in-memory computation are
unavailable to us at this point in time.  To estimate GRIM-3D's execution time,
we need to add up the time spent by three components (which we denote as $t_x$
for component $x$):}
\begin{itemize}
\item \chII{$t_1$: the time spent on read mapping,}
\item \chII{$t_2$: the time spent on coordinating which bins are examined by GRIM-Filter, and}
\item \chII{$t_3$: the time spent on applying the filter to each seed.}
\end{itemize}
\chII{To obtain $t_1$ and $t_2$, we measure the performance of \emph{GRIM-Software},
a software-only version of GRIM-Filter that does \emph{not} take advantage of 
processing in 3D-stacked memory.
We run GRIM-Software with mrFAST, and measure:}
\begin{itemize}
\item \chII{\emph{GRIM-Software-End-to-End-Time}, the end-to-end execution time for read mapping 
using GRIM-Software;}
\item \chII{\emph{GRIM-Software-Filtering-Time}, the time spent only on applying the filter (i.e., the GRIM-Filter portions of the code shown in Figure~\ref{fig:GRIM_integration}) 
using GRIM-Software.}
\end{itemize}
\chII{The values of $t_1$ and $t_2$ are the same for GRIM-Software and GRIM-3D, and we
can compute those by subtracting out the time spent on filtering from the end-to-end 
execution time: $t_1 + t_2 = \text{GRIM-Software-End-to-End-Time} - \text{GRIM-Software-Filtering-Time}$.
}
\chIII{
To estimate $t_3$, we use a validated simulator similar to
Ramulator~\cite{kim2015ramulator, ramulator_source}, which provides us with the
time spent by GRIM-3D on filtering using processing-in-memory. \chIII{The simulator
models the time spent by the in-memory logic to produce a seed location filter bitmask, and to store the
bitmask into a buffer that is accessible by the read mapper.}}
}

\paratitle{Evaluation System}
We evaluate the \chII{software versions of the read mappers 
(i.e., \chIV{mrFAST with FastHASH} and GRIM-Software)} using an Intel(R) Core
i7-2600 CPU running at 3.40GHz~\cite{intel_core}, with 16GB of DRAM for all experiments.

\paratitle{Data Sets}
\chI{We used ten real data sets from the 1000 Genomes Project~\cite{1000GP2012}. \chII{We used} the same data sets used by Xin et
al.~\cite{xin2013accelerating} \chII{for the original evaluation of mrFAST with FastHASH}, 
in order to provide a fair comparison \chII{to our baseline}.  Table~\ref{Tab:benchmark} lists the read
length and size of each data set.} 

\begin{table*}[!htbp]\centering
\footnotesize 
{\begin{tabular}{@{}lcllllll@{}}\toprule
&& ERR240726\_1 & ERR240727\_1 & ERR240728\_1 & ERR240729\_1 & ERR240730\_1\\\hdashline
No. of Reads && 4031354 & 4082203 & 3894290 & 4013341 & 4082472 \\
Read Length && 100 & 100 & 100 & 100 & 100\\\toprule
&& ERR240726\_2 & ERR240727\_2 & ERR240728\_2 & ERR240729\_2 & ERR240730\_2\\\hdashline
No. of Reads && 4389429 & 4013341 & 4013341 & 4082472 & 4082472 \\
Read Length && 100 & 100 & 100 & 100 & 100\\\toprule
&&&&&&
\end{tabular}}{}
\centering\caption{Benchmark data, obtained from the 1000 Genomes Project~\cite{1000GP2012}}
\label{Tab:benchmark}
\end{table*}

\paratitle{Code Availability}
\chIII{The code for GRIM-Filter, GRIM-Software, and our
simulator for 3D-stacked DRAM with processing-in-memory is freely available at} \\
\href{https://github.com/CMU-SAFARI/GRIM}{https://github.com/CMU-SAFARI/GRIM}.



\section{Evaluation Results}
\label{sec:results-analysis} 

\chI{We first profile the reference human genome in order to 1) determine a range
of parameters that are reasonable to use for GRIM-Filter.  We determine the
points of diminishing returns for several parameter values. This data is
presented in Section~\ref{subsec:motivational-data}. Using this preliminary
data, we reduce the required experiments to a reasonable range of parameters.
Our implementation of GRIM-Filter enables the variation of runtime parameters
(number of bins, token size, error tolerance, etc.) within
the ranges of values that we determine from our experimentation for the best
possible results.  We then quantitatively evaluate GRIM-Filter's improvement in
false negative rate and \chII{mapper} runtime over the baseline mrFAST with
FastHASH \chII{(Section~\ref{subsec:data})}.

\subsection{Sensitivity to GRIM-Filter Parameters} 
\label{subsec:motivational-data} 

In order to determine a range for the parameters for our experiments, we ran a
series of analyses on the fundamental characteristics of the human reference
genome. \chI{We perform these initial experiments to \chIII{1)} determine effective
parameters for \chII{GRIM-Filter \chIII{and 2)} compute its \emph{memory
footprint}}.  The memory footprint of GRIM-Filter depends directly on the
number of bins that we divide the reference genome into, since each bin
requires a bitvector to hold the token existence bits.  Since the bitvector
must contain a Boolean entry for each permutation of the token of size $n$,
each bitvector must contain $4^n$ bits. The total memory footprint is then
obtained by multiplying the bitvector size by the number of bins. In this
section, we sweep the number of bins, token size, and error tolerance of
GRIM-Filter while considering the memory footprint. To understand how each of
the different parameters affect the performance of GRIM-Filter, we study a
sweep on the parameters with a range of values that \chII{result in a memory
footprint under 16~GB (which is the current capacity of HBM2 on state-of-the-art
devices~\cite{NVIDIA-P100}).}} 

\paratitle{Average Read Existence}
Figure~\ref{plot:substr4} shows how varying a
number of different parameters affects the \emph{average read existence} across
the bins. We define average read existence to be the ratio of \chIII{bins with seed locations} that pass
the filter over all bins comprising the genome, for a representative set of
reads. We would like this value to be as low as possible because it reflects
the filter's ability to filter incorrect mappings. A lower average read
existence means that fewer bins must be checked when mapping the representative
set of reads. Across the three plots, we vary the token size from 4 to 6.
Within each plot, we vary the number of bins to split the reference genome
into, denoted by the different \chII{curves (with different colors and
markers)}.  The X-axis shows the \chI{error tolerance
\chII{that is} used}, and the Y-axis shows average read existence. We plot the
average and min/max across our 10 data sets (Table~1) as indicated,
respectively, by the triangle and whiskers.  

\begin{figure}[h]
    \centering
    \includegraphics[width=0.95\linewidth]{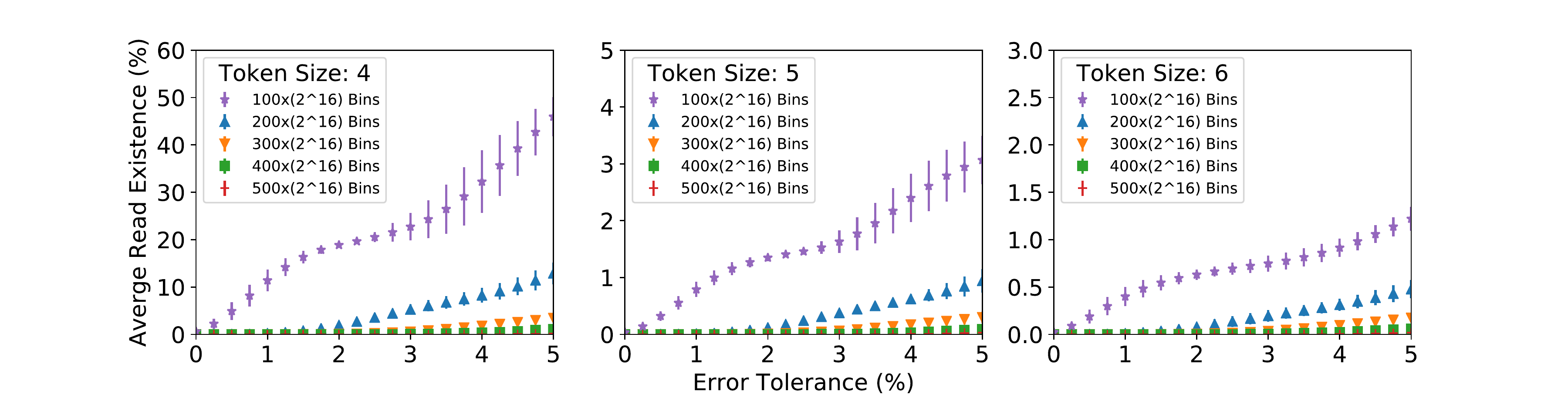}
    \caption{\chI{Effect of varying token size, error tolerance, and bin count on average read existence. We use a representative set of reads to collect this data. A lower value of average read existence represents \chIII{a more effective filter}. \chII{Note that the scale of the Y-axis is different for the three different graphs.}}} 
    \label{plot:substr4}
\end{figure}

We make \emph{three} observations from the figure.  First, looking across the
three plots, we observe that increasing the \emph{token size} from 4 to 5
provides a large \chII{(i.e., around 10x)} \chIII{reduction} in average read
existence, while increasing the token size from 5 to 6 \chIII{provides} a much
smaller \chII{(i.e., around 2x)} \chIII{reduction} in average read existence.
\chII{The reduction in average read existence is due to the fact that, in a
random pool of As, Cs, Ts, and Gs, the probability of observing a certain
substring of size $q$ is $({\frac{1}{4}})^q$. \chIII{Because} the distribution
of base pairs across a reference genome and across a bin is \emph{not} random,}
a larger token size does \emph{not} always result in a large decrease, \chII{as
seen when changing the token size from 5 to 6.  We note that increasing the
token size by one \chIV{causes GRIM-Filter to use 4x the memory footprint.}}
Second, we observe that in \chII{all three plots} (i.e., for all token sizes),
an increase in the number of bins results in a decrease in the \chII{average}
read existence.  This is because the bin size decreases as the number of bins
increases, and for smaller bins, we have a smaller sample size of the reference
genome that any given substring could exist within.\footnote{\chII{When
sweeping the number of bins, we use multiples of $2^{16}$ because
\chIII{$2^{16}$} is an even multiple of the number of TSVs between the logic
and memory layers in today's 3D-stacked memories (today's systems typically
have 4096 TSVs). \chIII{We want to use a multiple of $2^{16}$ so that we can
utilize all TSVs each time we copy data from a row buffer in the memory layer
to the corresponding row data register in the logic layer. \chV{This maximizes
GRIM-Filter's internal memory bandwidth utilization within 3D-stacked memory.}}}} 
Third, we observe that
for each plot, increasing the error tolerance results in an increase in the
\chII{average read existence. This is due to the fact that if we allow
more errors, fewer tokens of the entire read sequence must be present in a bin
\chIII{for a seed location \chIV{from that} bin} to pass the filter. This increases the
probability that a seed location of a random read passes the filter for a
random bin. \chIII{A poor sequence alignment at a location that passes the
filter is categorized as a false negative.}} \chII{We conclude from this figure
that using tokens of size 5 provides quite good filtering \chIII{effectiveness
(as measured by average read existence)} without requiring as much memory
footprint as \chIV{using a token size of 6}.}

\paratitle{False Negative Rate}
We choose our final bitvector size after sweeping
the number of bins and the error tolerance ($e$).
Figure~\ref{plot:buckets_FPR} shows how varying these parameters affects the
false negative rate of GRIM-Filter. The X-axis varies the number
of bins, while the different lines represent different values of $e$.

\begin{figure}[h]
    \centering
    \includegraphics[width=0.5\linewidth]{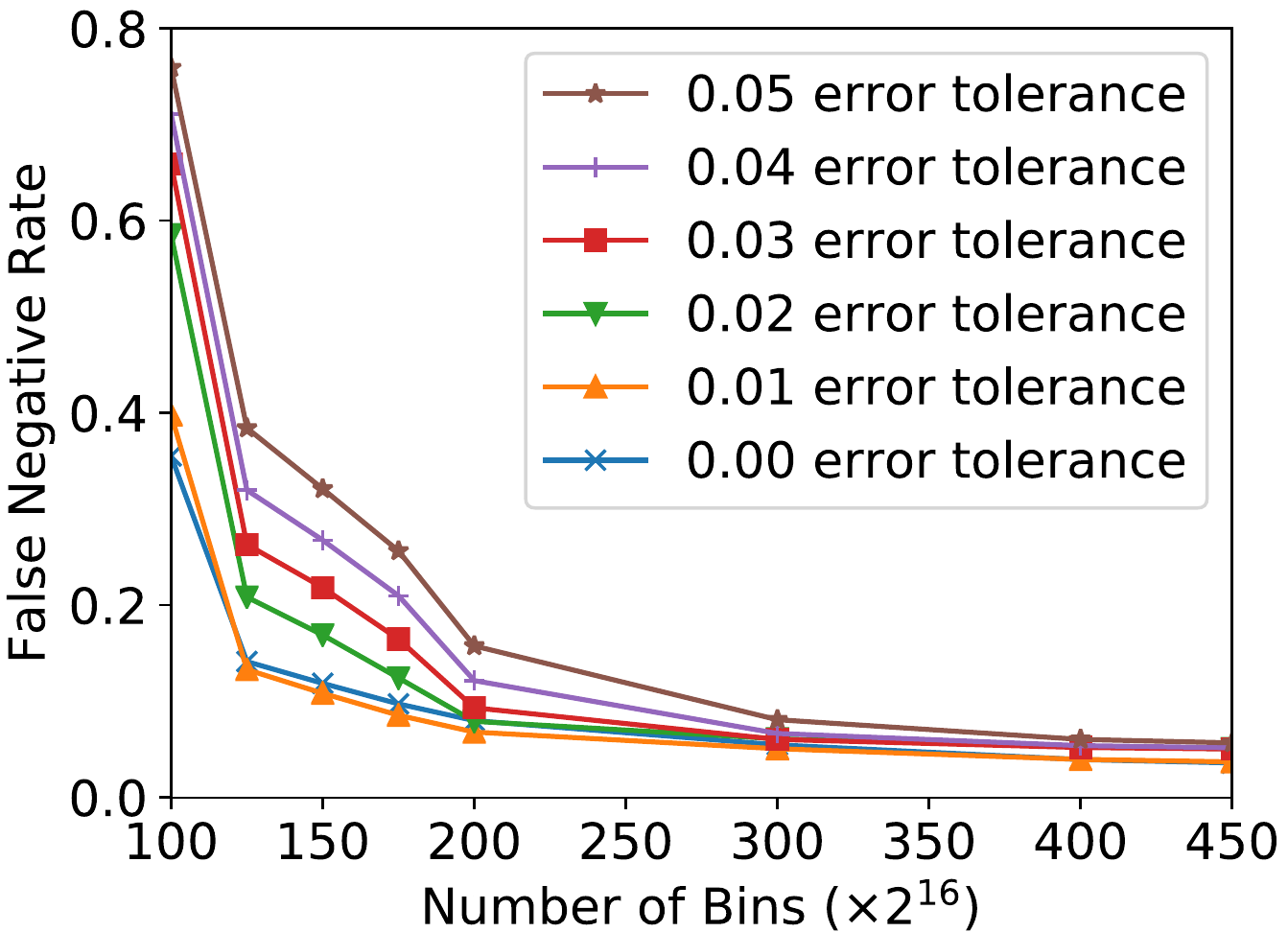}
    \caption{\chII{GRIM-Filter's false negative rate (lower is better) as we vary the number of bins. We find that increasing the number of bins 
beyond 300$\times$$2^{16}$ yields diminishing improvements in the false negative rate, regardless of the error tolerance value.}}
    \label{plot:buckets_FPR}
\end{figure}

We make two observations from this figure.  First, we find that, with more bins
(i.e., with a smaller bin size), the false negative rate (i.e., the
\chII{fraction} of locations that pass the filter, but do not result in a
mapping after alignment) decays exponentially. \chIII{Above 300$\times$$2^{16}$ bins,
we begin to see diminishing returns on the reduction in false negatives for all
error tolerance values}. Second, we observe that, as we increase the error
tolerance, regardless of the other parameters, the false negative rate
increases. \chIII{We also find that} the number of bins \chIII{1)} minimally
affects the runtime of GRIM-Filter (not plotted) and \chIII{2)} linearly
increases the memory footprint. \chIII{Based on this study}, we choose to use
450$\times2^{16}$ bins, which reflects a reasonable memory footprint (see
below) with the other parameters.}

\paratitle{Memory Footprint}
\chI{A larger number of bins results in more
bitvectors, so we must keep this parameter at a reasonable value in order to
retain a reasonable memory footprint for GRIM-Filter. Since we have chosen a
token size of 5, GRIM-Filter requires $t$ bitvectors with a length of
$4^5=1024$, where $t$ equals the number of bins we segment the reference genome
into.  We conclude that employing 450$\times$$2^{16}$ bins results in the best
trade-off between memory footprint, filtering efficiency, and
runtime.\footnote{We note that the time to generate the bitvectors is
\emph{not} included in our final runtime results, because these need to be
generated only once per \chIII{reference} genome, either by the user or by the
distributor. \chIII{We} find that, with a genome of length $L$, we can generate the
bitvectors in $(9.03e-08) \times L$ seconds when we use 450$\times$$2^{16}$
bins (this is approximately 5~minutes for the human genome).}  This set of
parameters results in a total memory footprint of approximately $3.8$ GB for
storing the bitvectors of this mechanism, which is a very reasonable size for
today's 3D-stacked memories\chII{~\cite{AMD-R9-Graphics, AMD-RX-Vega-Graphics,
NVIDIA-P100, altera-HMC-UG, HMC-sources, standard2013high, AMD-HBM}.}
}

\paratitle{GRIM-Filter Parallelization}
\chI{GRIM-Filter operates on every
bin independently and in parallel, \chII{using a separate logic module for each
bin}. Thus, GRIM-Filter's parallelism increases with each additional bin it
operates on simultaneously.  \chII{We refer to the \chIII{set of consecutive}
bins that the GRIM-Filter logic modules are currently assigned to as the
\emph{bin window} ($w$). The internal bandwidth of HBM2~\cite{o2014highlights}
enables copying 4096 bits from a memory layer to the logic layer every cycle,
allowing GRIM-Filter to operate on as many as 4096~consecutive bins in parallel
(i.e., it has a bin window of size $w=4096$).} \chIII{GRIM-Filter must only
check bin windows that contain at least one seed location (i.e., a span of 4096
consecutive bins with \emph{zero} seed locations does not need to be checked).
In contrast, if a consecutive set of 4096 bins contains many seed locations,
GRIM-Filter can operate on every bin in parallel and quickly determine which
seed locations within the 4096 bins can safely be discarded. In these cases,
GRIM-Filter can most effectively utilize the parallelism available from the
4096~independent logic modules.}

\chII{\chIII{In order to understand GRIM-Filter's \chIV{ability to parallelize
operations on many bins}}, we analyze GRIM-Filter when using a bin window of size
$w=4096$, which takes advantage of the full memory bandwidth available in HBM2
memory. 
\chV{As we discuss in Section~\ref{sec:filter:integration}, the read mapper
generates a list of potential seed locations for a read sequence,
and sends this list to GRIM-Filter when the filter starts.  Several bins, which we call
\emph{empty bins}, do not contain \emph{any} potential seed locations.}
When $w=1$, there is only one
logic module, and \chV{if the module is assigned to an empty bin,}
GRIM-Filter immediately moves on to the
next bin \chV{without computing the accumulation sum}. 
\chV{However, when $w=4096$, some, but not all, of the logic modules may be
assigned to empty bins.  This happens because in order to simplify the hardware,
GRIM-Filter operates all of the logic modules in lockstep
(i.e., the filter fetches a \emph{single row} from each bank of memory, which
includes the existence bits \emph{for a single token} across multiple rows, and
all of the logic modules read and process the existence bits
for the same token in the same cycle).
Thus, a logic module assigned to an empty bin must wait for the
other logic modules to finish before it can move onto another bin.
As a result, GRIM-Filter with $w=4096$ is not 4096x faster than
GRIM-Filter with $w=1$.
To quantify the benefits of parallelization, we compare the performance of 
GRIM-Filter with these two bin window sizes
using a representative set of reads.}
For 10\% of the seeds, we find that GRIM-Filter with $w=4096$
reduces the filtering time by 98.6\%, compared to GRIM-Filter with $w=1$.  For
the remaining seeds, we find that GRIM-Filter with $w=4096$ reduces the
filtering time by 10--20\%.  Thus, even though many of the logic modules 
\chV{are assigned to empty bins in a given cycle}, GRIM-Filter reduces the filtering time 
by operating on many bins \chV{that contain potential seed locations} in parallel.}

\paratitle{\chV{Overlapping GRIM-Filter Computation with Sequence Alignment in the CPU}}
\chII{In addition to operating on multiple bins in parallel, one benefit of
implementing GRIM-Filter in 3D-stacked memory is that filtering operations can
be parallelized with sequence alignment \chV{that happens on the CPU}, since filtering no longer uses the
CPU.  Every cycle, for a bin window of size $w=4096$, \chV{GRIM-Filter's Filter
Bitmask Generator (\incircle{1} in Figure~\ref{fig:GRIM_integration})} reads
4096~bits from memory, and updates the accumulation sums for the bins within
the bin window that contain a potential seed location. Once the \chVI{accumulation
sums are} computed and compared against the threshold, \chV{GRIM-Filter's
Seed Location Checker (\incircle{2} in Figure~\ref{fig:GRIM_integration})} can discard
seed locations that map to bins whose accumulation sums do \chVI{\emph{not}} meet the
threshold (i.e., the seed locations that should not be sent to sequence
alignment). \chIII{The seed locations that are not discarded \chV{are sent} to the
read mapper for sequence alignment \chV{(\incircle{4} in Figure~\ref{fig:GRIM_integration})}}, ending GRIM-Filter's work for the current bin
window.  While the read mapper aligns the sequences that passed through the
filter from the completed bin window, \chV{GRIM-Filter's Filter Bitmask Generator} moves onto another bin
window, computing the seed location filter bits for a new set of bins.  If
GRIM-Filter can exploit enough parallelism, it can provide \chV{the CPU}
with enough bins to keep the \chV{sequence} alignment step busy for at least as long as the
time needed for \chV{the Filter Bitmask Generator} to process the new bin window.  This would allow
the filtering latency to overlap \chVI{\emph{completely}} with alignment, in effect hiding
\chVI{GRIM-Filter's} latency.  We find that a bin window of 4096~bins provides enough
parallelism to \chV{\emph{completely}} hide the \chV{filtering latency} while the read mapper
\chV{running on the CPU performs sequence alignment}.}

\subsection{Full Mapper Results} 
\label{subsec:data} 

We use a popular seed-and-extend mapper, mrFAST~\cite{alkan2009personalized},
to retrieve all candidate mappings from the ten real data sets we evaluate (see
Section~\ref{sec:exp_method}). In our experiments, we use a token size of 5 and
450$\times$$2^{16}$ bins, as discussed in
Section~\ref{subsec:motivational-data}. All remaining parameters specific to
mrFAST are held at the default values across all of our evaluated read mappers. 

\paratitle{False Negative Rate}
Figure~\ref{plot:FPR} shows the false negative rate
of GRIM-Filter compared to the baseline \chII{FastHASH filter} across the ten real
data sets we evaluate. The six plots in the figure show false negative rates
for \chII{error tolerance values (i.e., $e$) ranging from 0.00 to 0.05, in
increments of 0.01}.  \chI{We make three observations from the figure. First,
GRIM-Filter provides a much lower false negative rate than the baseline
\chII{FastHASH filter} for all data sets and for all error tolerance values.
For an error tolerance of $e=0.05$ (shown in the bottom graph),\footnote{An 
error tolerance of $e=0.05$ is widely used in alignment during DNA read
mapping~\cite{ahmadi2012hobbes, cheng2015bitmapper, hatem2013benchmarking,
xin2015shifted}.} the false negative rate for GRIM-Filter is \emph{5.97x} lower
than for FastHASH filter, averaged across all 10~read data sets. Second, GRIM-Filter's
false negative rate 1)~increases as the error tolerance increases from
\chII{$e=0.00$ to $e=0.02$}, and then 2)~decreases as the error tolerance
increases further from \chII{$e=0.03$ to $e=0.05$}.  There are at least two
conflicting reasons. First, as the error tolerance increases, the accumulation
sum threshold decreases \chII{(as shown in Figure~\ref{fig:Threshold_eq})}} and
thus GRIM-Filter discards \emph{fewer} locations, which results in a
\emph{higher} false negative rate. Second, as the error tolerance increases,
the number of acceptable \chIII{(i.e., correct)} mapping locations increases
while the number of candidate locations remains the same, which results in a
\emph{lower} false negative rate.  The interaction of these conflicting
reasons results in the initial increase and the subsequent decrease in the
false negative rates that we observe.  Third, we observe that for higher
\chII{error tolerance values}, GRIM-Filter reduces the false negative rate
compared to \chII{the FastHASH filter} by a larger fraction.  This shows
that GRIM-Filter is much more effective at filtering mapping locations when
we increase the error tolerance.  \chII{We conclude that GRIM-Filter is
very effective in reducing the false negative rate.}

\begin{figure}[h]
    \centering
    \includegraphics[width=0.95\linewidth]{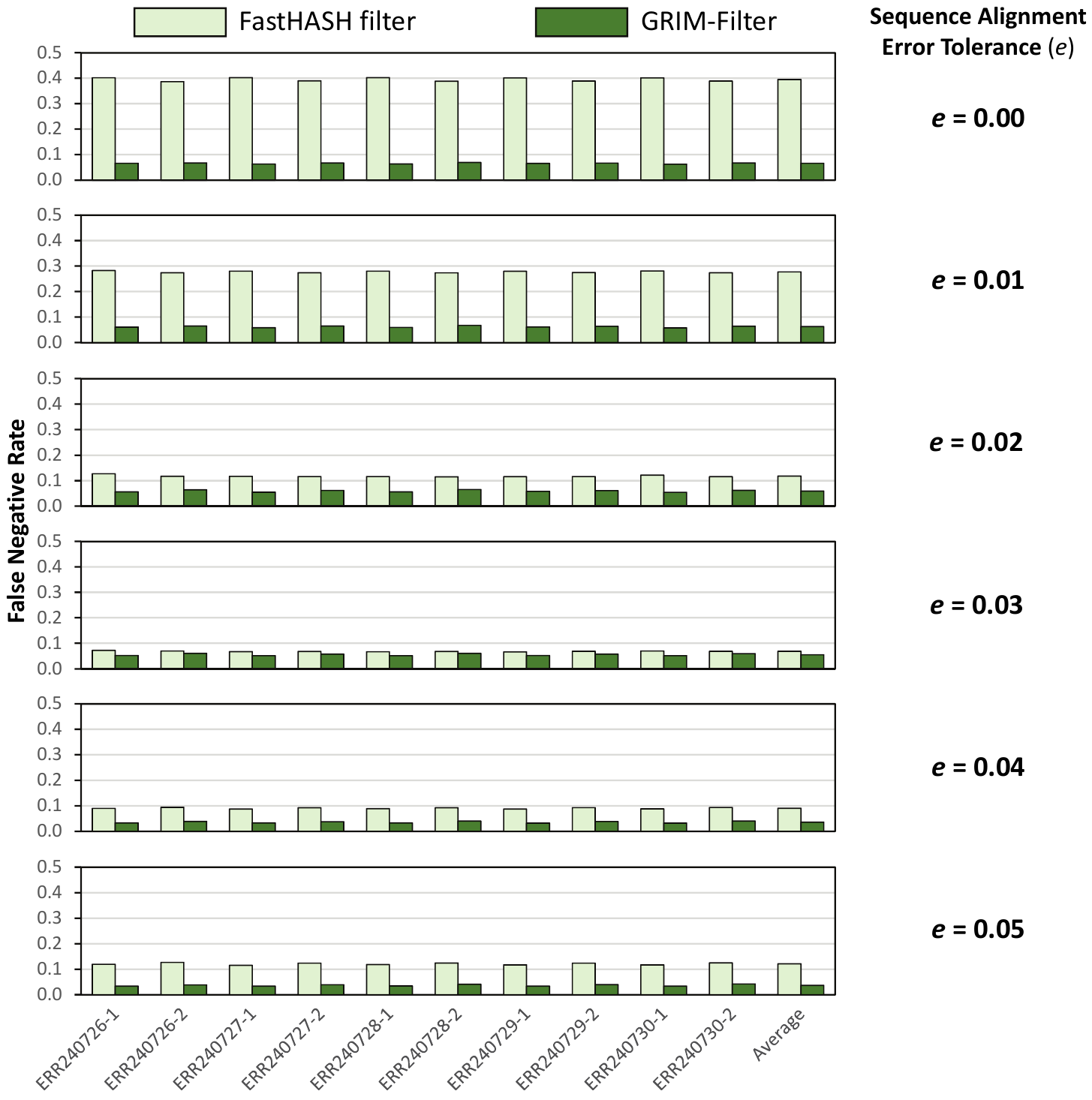} 
    \caption{False negative rates of GRIM-Filter and FastHASH filter  
    across ten real data sets for six different error tolerance values.}
    \label{plot:FPR}
\end{figure}

\paratitle{Execution Time}
\chI{Figure~\ref{plot:runtime} compares the execution
time of GRIM-3D to that of mrFAST with FastHASH across all ten different read
data sets for the same error tolerance values used in Figure~\ref{plot:FPR}.
We make three observations.} \chI{First, \chII{GRIM-3D} improves performance for
all of our data sets for all error tolerance values. For an error tolerance of
$e=0.05$, the average (maximum) performance improvement is 2.08x (3.65x) across all
10 data sets. Second, as the error tolerance increases, GRIM-3D's
performance improvement also increases.  This is because \chVI{GRIM-Filter} safely
discards many more mapping locations than \chVI{the} FastHASH filter at higher error tolerance
values (as we showed in Figure~\ref{plot:FPR}).  Thus, \chVI{GRIM-Filter} saves
significantly more execution time than \chVI{the FastHASH filter} by ignoring many more
unnecessary alignments. Third, based on an analysis of the execution time
breakdown of GRIM-3D (not shown), we find that GRIM-3D's performance gains are
mainly due to an 83.7\% reduction in the average computation time spent on
false negatives, compared to using the FastHASH filter for seed location filtering.
We conclude that employing GRIM-Filter for seed location filtering in a
state-of-the-art read mapper significantly improves the performance of the read
mapper.} 

\begin{figure}[h]
    \centering
    \includegraphics[width=0.95\linewidth]{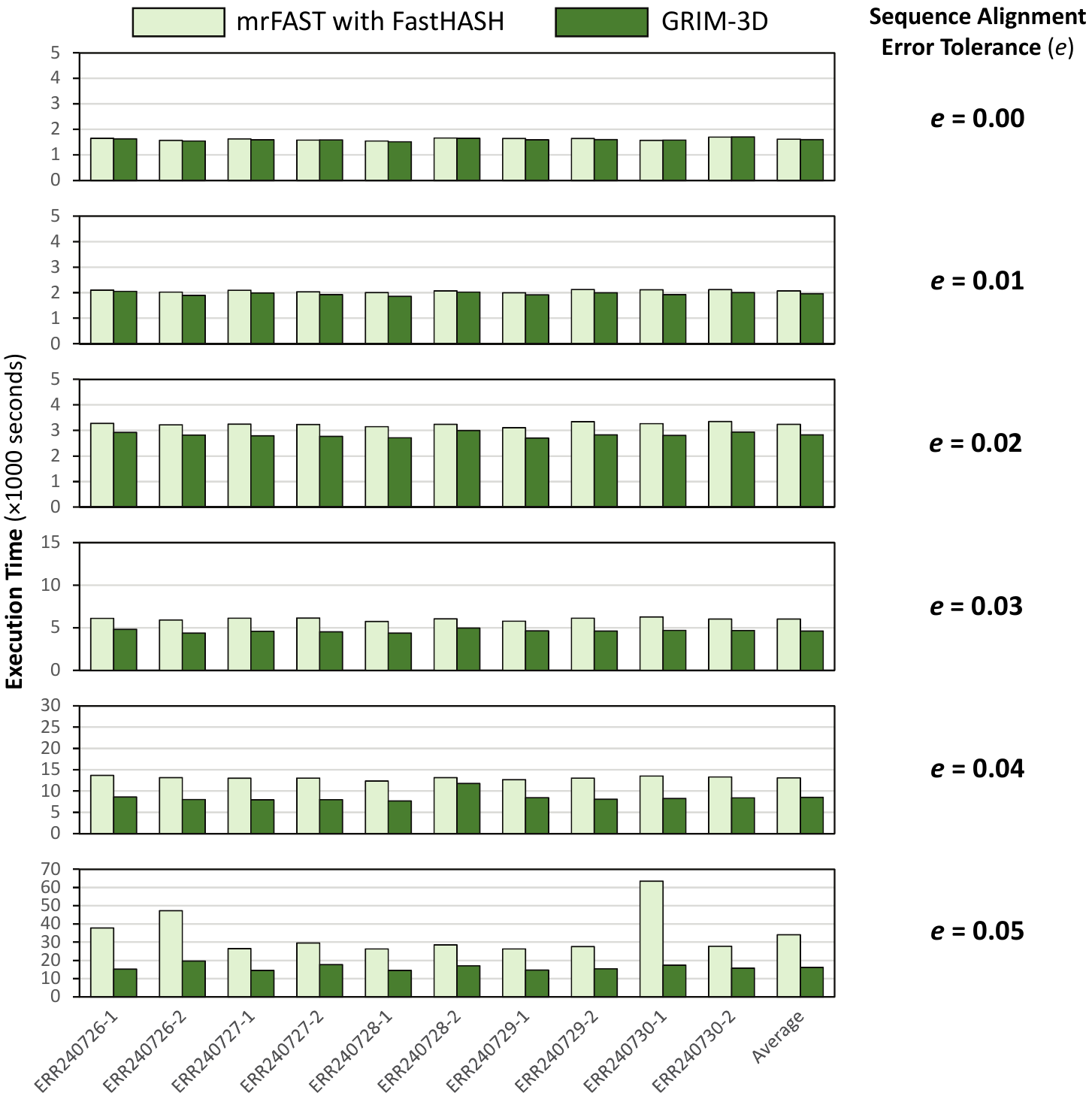} 
    \caption{\chII{Execution time of two mappers, GRIM-3D and \chIV{mrFAST with FastHASH}, across ten real data sets for six different error tolerance values.}
    \chII{Note that the scale of the Y-axis is different for the six different graphs.}} 
    \label{plot:runtime}
\end{figure}

\section{Related Work} 
\label{sec:related} 

\chI{To our knowledge, this is the first paper to exploit 3D-stacked DRAM and
its processing-in-memory capabilities to implement a new seed location
filtering algorithm that mitigates the major bottleneck in read mapping,
pre-alignment \chII{(i.e., seed location filtering)}.  In this section, we
briefly describe related works that aim to 1)~accelerate pre-alignment
algorithms, and 2)~accelerate \chII{sequence} alignment with hardware support.} 

\paratitle{Accelerating Pre-Alignment}
\chI{A very recent prior work~\cite{alser2016gatekeeper}} implements a seed location filter in an FPGA,
and shows significant speedup against prior filters. 
\chII{However, as shown in that work, the FPGA is still limited by the
memory bandwidth bottleneck.  GRIM-Filter can overcome this bottleneck
on an FPGA as well.}


\paratitle{Accelerating Sequence Alignment}
\chI{\chIII{Another} very recent prior work~\cite{liu20173d}
exploits the high memory bandwidth and the reconfigurable logic layer of
3D-stacked memory to implement an accelerator for sequence alignment (among
other basic algorithms within the sequence analysis pipeline).  Many prior
works (e.g., \cite{aluru2014review, arram2013hardware, arram2013reconfigurable,
ashley2010clinical, chiang2006hardware, hasan2007hardware, houtgast2015fpga,
mcmahon2008accelerating, olson2012hardware, papadopoulos2013fpga,
waidyasooriya2014fpga}) \chIII{use} FPGAs to also accelerate \chII{sequence}
alignment. These works accelerate sequence alignment using customized FPGA
implementations of different existing read mapping algorithms. For example, Arram et
al.~\cite{arram2013reconfigurable} accelerate the SOAP3 tool on an FPGA engine,
achieving up to 134x speedup compared to BWA~\cite{li2010fast}. 
Houtgast et al.~\cite{houtgast2015fpga} present an FPGA-accelerated version of
BWA-MEM that is 3x faster compared to its software implementation.
Other works use GPUs~\cite{blom2011exact, liu2012soap3, luo2013soap3,
manavski2008cuda} \chII{for the same purpose of accelerating sequence
alignment. For example,} Liu et al.~\cite{liu2012soap3} accelerate BWA and
Bowtie by 7.5x  and  20x, respectively.  In contrast to
GRIM-Filter, all of these accelerators focus on accelerating \chII{sequence}
alignment, whereas GRIM-Filter accelerates pre-alignment (i.e., seed location
filtering). Hence, GRIM-Filter is orthogonal to these works, and can be
combined with any of them for further performance improvement.}



\section{Future Work} 
\label{sec:future_work}

We have shown that GRIM-Filter significantly \chIII{reduces} the execution time of
read mappers by reducing the number of unnecessary sequence alignments
and by taking advantage of processing-in-memory using 3D-stacked DRAM technology.
We believe there are many other possible applications for employing 3D-stacked DRAM
technology within the genome sequence analysis pipeline (as initially explored
in \cite{liu20173d}), and significant additional performance \chII{improvements
can be obtained} by combining future techniques with GRIM-Filter. Because
GRIM-Filter is essentially a seed location filter to be employed before sequence 
alignment during read mapping, it can be used in any other read mapper along
with any other acceleration mechanisms in the genome sequence analysis pipeline.

We identify three promising major future research directions. We believe
it is promising to 1)~explore the benefits of combining
GRIM-Filter with other various read mappers in the field, 2)~show the effects
of mapping to varying sizes of reference genomes, and 3)~examine how GRIM-Filter
can scale to process a greater number of reads concurrently.


\section{Conclusion} 

\chI{This paper introduces GRIM-Filter, a novel algorithm for seed
location filtering, which is a critical performance bottleneck in genome read
mapping.  \chIII{GRIM-Filter has three major novel aspects. First,} it preprocesses the reference genome to collect
metadata on large subsequences (i.e., \emph{bins}) of the genome} and stores
information on whether small subsequences (i.e., \emph{tokens}) are present in
each bin. \chIII{Second,} GRIM-Filter efficiently operates on the metadata to quickly determine
whether to discard a mapping location for a read sequence prior to an expensive
sequence alignment, thereby reducing the number of unnecessary alignments and
improving performance. \chIII{Third,} GRIM-Filter takes advantage of the logic layer within
3D-stacked memory, which enables the efficient use of processing-in-memory to
overcome the memory bandwidth bottleneck in seed location filtering. We examine
the trade-offs for various parameters in GRIM-Filter, and present a set of
parameters that result in significant performance improvement over the
state-of-the-art seed location filter, FastHASH. \chII{When running with a
sequence alignment error tolerance of 0.05, we show that GRIM-Filter 1)~filters
seed locations with \emph{5.59x--6.41x} lower false negative rates than
FastHASH; and 2)~improves the performance of the fastest read mapper, mrFAST
with FastHASH, by \emph{1.81x--3.65x}.}  GRIM-Filter is a universal \chIII{seed location} filter that
can be applied to any read mapper.}

\chII{We believe there is a very promising potential in \chV{designing} DNA read
mapping algorithms \chV{for} \chIII{new memory technologies (like 3D-stacked DRAM)} and new
processing paradigms (like processing-in-memory).} We hope that the results from
our paper \chIII{provides inspiration for other works to design new sequence
analysis and other bioinformatics algorithms that take advantage of new memory technologies and new processing paradigms, such as processing-in-memory using 3D-stacked DRAM.}

\begin{backmatter}

%

\section*{Acknowledgments}
  An earlier version of this paper appears on arXiv.org~\cite{kim2017grim}. An
  earlier version of this work was presented as a short talk at
  RECOMB-Seq~\cite{kim2016grimRECOMB}. We thank the anonymous reviewers for
  feedback. This work was supported in part by the Semiconductor Research
  Corporation, \chIV{the National Institutes of Health (grant HG006004 to O. Mutlu and C. Alkan),} Intel, Samsung,
  and VMware. 


\bibliographystyle{bmc-mathphys} 
{\footnotesize\bibliography{references}}      

\end{backmatter}

\ifcameraready
\else
  \newpage
  \input{10_response}
\fi

\end{document}